\documentclass{egpubl}
\usepackage{eg2024}
 
\ConferencePaper        
\CGFStandardLicense

\usepackage[T1]{fontenc}
\usepackage{dfadobe}  

\usepackage{cite}  
\BibtexOrBiblatex
\electronicVersion
\PrintedOrElectronic
\ifpdf \usepackage[pdftex]{graphicx} \pdfcompresslevel=9
\else \usepackage[dvips]{graphicx} \fi

\usepackage{egweblnk} 

\usepackage{mathtools}
\usepackage{amsfonts}
\usepackage[inline]{enumitem}
\usepackage{algorithm}
\usepackage{algpseudocode}

\usepackage{amsthm}

\usepackage{cleveref}
\usepackage{nicefrac}
\usepackage{overpic}
\usepackage{tikz}

\def\Reals{\mathbb{R}}
\newcommand{\Realsn}[1]{\Reals^{#1}}
\newcommand{\manifold}[1]{\mathcal{#1}}
\newcommand{\mesh}[1]{\Hat{\manifold{#1}}}
\def\M{\manifold{M}}
\def\DiscM{\mesh{M}}
\def\N{\manifold{N}}
\def\DiscN{\mesh{N}}
\def\BPhi{\Phi}
\def\BPsi{\Psi}
\def\DBPhi{\Hat{\BPhi}}
\def\DBPsi{\Hat{\BPsi}}
\def\PtoP{\Pi}
\def\LRPtoP{\Hat{\PtoP}}

\def\BigOSym{\mathcal{O}}
\newcommand{\BigO}[1]{\BigOSym\left({#1}\right)}

\newcommand{\myvec}[1]{\boldsymbol{#1}}
\newcommand{\mymat}[1]{\mathbf{#1}}

\DeclareMathOperator*{\argmax}{arg\,max}

\def\etal{\emph{et al.}}

\usepackage{xcolor}

\definecolor{myred3}{RGB}{250,3,3} 
\definecolor{myred2}{RGB}{180,3,3} 
\definecolor{myred1}{RGB}{110,3,3} 
\definecolor{BlueMATLAB}{HTML}{0072BD}
\definecolor{GreenMATLAB}{HTML}{77AC30}

\newcommand{\TableFirst}[1]{{\color{BlueMATLAB} #1}}
\newcommand{\TableSecond}[1]{{\color{GreenMATLAB} #1}}

\newtheorem{thm}{Theorem}
\newtheorem{dfn}{Definition}
\newtheorem{prp}{Proposition}
\newenvironment{customthm}[1]
  {\renewcommand\thethm{#1}\thm}
  {\endthm}
\newenvironment{customprp}[1]
  {\renewcommand\theprp{#1}\prp}
  {\endprp}

\def\eg{\emph{e.g.}}
\def\ie{\emph{i.e.}}
\def\etal{\emph{et al.}}

\title[ReMatching: Low-Resolution Representations for Scalable Shape Correspondence]%
      {ReMatching: Low-Resolution Representations for Scalable Shape Correspondence}

\author[F. Maggioli \& D. Baieri \& E. Rodol\`{a} \& S. Melzi]
{\parbox{\textwidth}{\centering%
F. Maggioli$^{2}$\orcid{0000-0001-8008-8468}
and 
D. Baieri$^{1}$\orcid{0000-0002-0704-5960}
and
E. Rodol\`{a}$^{1}$\orcid{0000-0003-0091-7241}
and
S. Melzi$^{2}$\orcid{0000-0003-2790-9591}
}
\\
{\parbox{\textwidth}{\centering%
$^1$Sapienza - University of Rome, Italy\\
$^2$Universit\`{a} di Milano Bicocca, Italy
}
}
}

\begin{document}

\teaser{
 \includegraphics[width=\linewidth]{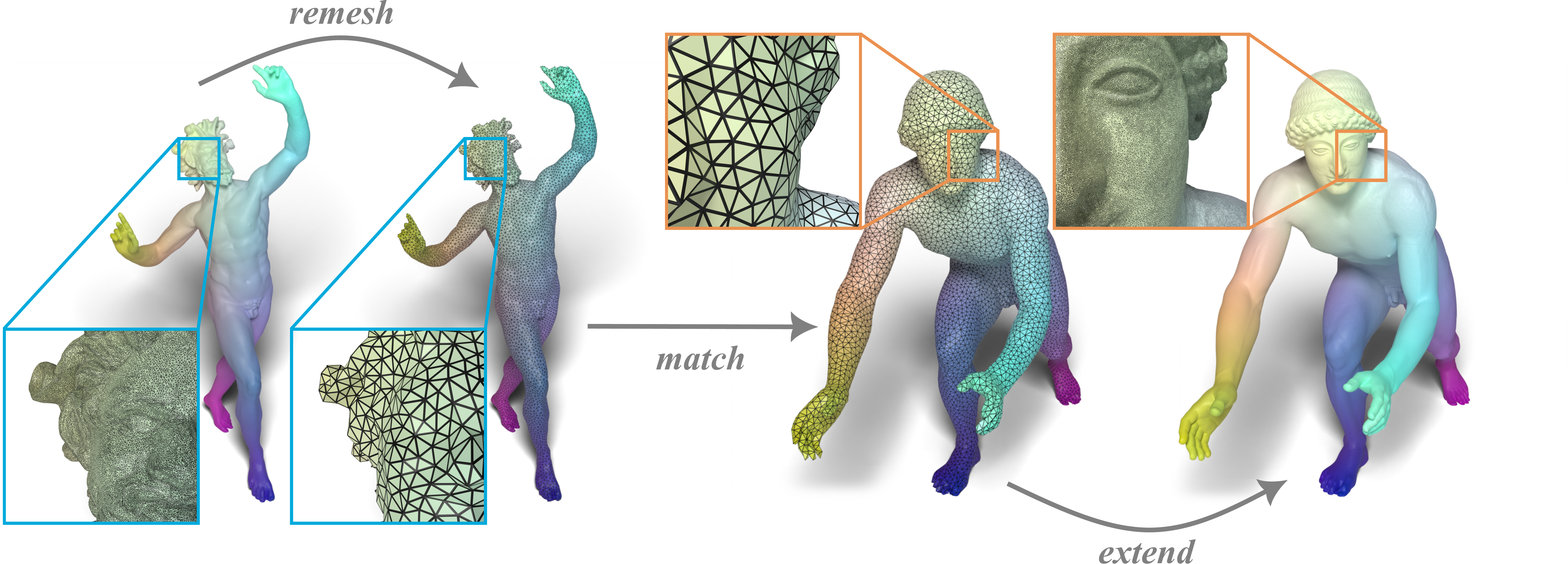}
 \centering
  \caption{Correspondence between two dense shapes computed with our method, using ZoomOut~\cite{melzi:2019:zoomout} on low-resolution meshes. The coordinates functions of the dancing faun statue on the left (\textasciitilde~750k vertices) are transferred using a functional map to the Aphaea warrior statue on the right (\textasciitilde~3.5M vertices). The colors encode the correspondence. Despite the mesh density, shown in the close-up, the computation took \textasciitilde~2 minutes.}
\label{fig:teaser}
}

\maketitle
\begin{abstract}

We introduce \emph{ReMatching}, a novel shape correspondence solution based on the functional maps framework. Our method, by exploiting a new and appropriate \emph{re}-meshing paradigm, can target shape-\emph{matching} tasks even on meshes counting millions of vertices, where the original functional maps does not apply or requires a massive computational cost. The core of our procedure is a time-efficient remeshing algorithm which constructs a low-resolution geometry while acting conservatively on the original topology and metric. These properties allow translating the functional maps optimization problem on the resulting low-resolution representation, thus enabling efficient computation of correspondences with functional map approaches. Finally, we propose an efficient technique for extending the estimated correspondence to the original meshes. We show that our method is more efficient and effective through quantitative and qualitative comparisons, outperforming state-of-the-art pipelines in quality and computational cost.

		
		

\end{abstract}  


\section{Introduction and related work}\label{sec:intro}

%

The task of finding a semantically meaningful correspondence between discrete surfaces has always been a fundamental topic in the field of shape analysis. Researchers developed a wealth of solutions for this application, and new methods continue to be designed~\cite{deng:2022:matchingsurvey,sahilliouglu:2020:matchingsurvey}.

Among the various approaches, the functional maps framework~\cite{ovsjanikov:2012:fmaps,ovsjanikov:2016:fmapsmatching} received significant attention.
Rather than finding a point-wise correspondence, the functional maps framework aims to define a correspondence between functions, encoding it into a small matrix. 
Following this direction, a variety of works tried to improve the actual computation of the map~\cite{donati:2020:deepfmaps,melzi:2019:zoomout,ren:2018:continuousfmaps} or to extend the framework to different types of bases~\cite{melzi:2018:localeigs,nogneng:2018:eigprods,maggioli:2021:orthoprods}.

Despite the amount of research outcomes based on the functional maps framework, both exploiting geometric properties and tools~\cite{eynard:2016:coupledfmaps,burghard:2017:embeddinggreen,ezuz:2017:denoisemaps} and clever optimization techniques~\cite{nogneng:2017:descriptorpreserve,ren:2020:maptree,ren:2021:discreteopt}, the problem of computing the map is still bounded by the time complexity of sparse eigendecompositions~\cite{kressner:2006:numerical}.

To overcome this problem, researchers started investigating scalable solutions to the Laplace-Beltrami eigenproblem. In this regard, multi-resolution techniques~\cite{vaxman:2010:multiresheat,nasikun:2018:fasteigs,shoham:2019:hierarchicalmaps,nasikun:2022:hierarchicaleigs} proved to be very effective for scalable computation of spectral quantities for tasks such as retrieval and mesh filtering~\cite{reuter:2006:shapedna,levy:2010:spectralprocessing}, but their applicability to the functional map framework proved to require additional care~\cite{magnet:2023:scalablefmaps}. 
Furthermore, recent research proposed spectral coarsening methods~\cite{lie:2019:spectralcoarse} or spectral preserving simplification techniques~\cite{lescoat:2020:spectraldecimation}. However, these techniques usually rely on the computation of the full-resolution spectrum, making them unsuitable for large-scale applications.

Closely related to our work is the contribution from Magnet \etal~\cite{magnet:2023:scalablefmaps}, which, for the first time, proposes a solution for scaling the functional maps approach to high-resolution meshes. In their work, the authors reduce the dimensionality of the eigenproblem, inducing a linear relationship between the functional space on the mesh and a lower-dimension functional space on a sparse sample. The extreme sparsity of the sampling, combined with its efficient computation, leads to an algorithm that can effectively deal with high-density meshes. Nevertheless, this type of reduction suffers from a major limitation: it does not mitigate the negative influence that small components and local details at high frequency could have on the alignment of the spectra.

Finally, our pipeline relies on building robust and sparse representations of dense meshes while still keeping a bijective correspondence between the high- and low-resolution shapes, and it is worth mentioning that other works have already attempted to do so. Jiang \etal~\cite{jiang:2020:bijectiveshell} propose building a lower resolution prismatic shell that preserves a bijectivity with the original surface. However, their algorithm does not scale well to meshes with a high triangle count, gives no control and no guarantees on the number of output vertices, and requires many assumptions on the input shape (\eg, manifoldness, orientability, no self-intersections). On the other hand, the solution proposed by Liu \etal~\cite{liu:2023:intrinsicsimplification} exploits an intrinsic error metric for decimating a mesh to a given size while keeping track of a geodesic barycentric mapping each triangle of the resulting shape to a geodesic triangle on the input surface. Despite the robustness of the method, the simplification approach makes it unsuitable for a functional maps setting, where it is required to decimate shapes from millions of vertices to a few thousand. We will further discuss this limitation in \Cref{sec:results:performance}.

With our work, we introduce a new scalable functional map pipeline that, while efficiently handling meshes with high vertex density (see \Cref{fig:teaser}), yields stable results and is not bounded by the quality of the original triangulation. To summarize our contribution:
\begin{itemize}[topsep=0pt,noitemsep]
    
    \item we translate the matching pipeline to low-resolution representations, enabling a fast and scalable computation of functional maps between dense shapes, providing the first alternative to~\cite{magnet:2023:scalablefmaps};
    
    \item we propose a new efficient and geometry-preserving remeshing algorithm specifically designed for our pipeline;

    \item we exploit a fast solution for extending scalar maps from the low-resolution remeshed shape to the original surface, thus efficiently obtaining suitable bases for function transfer and shape matching.
    

\end{itemize}

\section{Background}\label{sec:background}

In the discrete setting, we represent a shape as a triangular mesh $\M = (V, E, T)$, where
\begin{enumerate*}[noitemsep,topsep=0pt,label={(\roman*)}]
    \item $V \subset \Reals^3$ is a set of vertices;
    \item $E \subset V^2$ is a set of edges between vertices;
    \item $T \subset V^3$ is a set of triangles composing the surface.
\end{enumerate*}

\subsection{Functional maps}

We discretize a scalar function $f : \M \to \Reals$ as a signal defined on the vertices $V$ and represented as a vector $\myvec{f} \in \Reals^{|V|}$. Hence, the Laplace-Beltrami operator $\Delta_{\M} : \mathcal{F}(\M) \to \mathcal{F}(\M)$ is discretized as a sparse matrix $\mymat{L}_{\M} \in \Reals^{|V|\times|V|}$, generally represented by means of a stiffness matrix $\mymat{S}_{\M}$ and a mass matrix $\mymat{A}_{\M}$~\cite{pinkall93,meyer03}. The eigendecomposition $\mymat{S}_{\M} \BPhi_{\M} = \mymat{A}_{\M} \BPhi_{\M} \Lambda_{\M}$ of the Laplacian yields an orthonormal basis for the functional space on the surface~\cite{levy06laplace,levy:2010:spectralprocessing}. This basis has some analogies with the Fourier basis and is optimal to represent smooth functions when the basis is truncated~\cite{aflalo:2015:optimallaplacian}.

Given two shapes $\M$ and $\N$, respectively with $m$ and $n$ vertices, a point-to-point map $\PtoP : \M \to \N$ can be expressed as a binary matrix $\PtoP \in \{ 0, 1 \}^{m \times n}$ such that $\PtoP(i,j) = 1$ if the vertex $j$-th of $\N$ corresponds to the vertex $i$-th of $\M$, and $\PtoP(i,j) = 0$ otherwise. 
Any point-wise correspondence $\PtoP$ estabilishes a functional map $T_{\PtoP} : \mathcal{F}(\N) \to \mathcal{F}(\M)$ as $T_{\PtoP}(f) = f \circ \PtoP$, $\forall f \in \mathcal{F}(\N)$. 
%



As proposed in~\cite{ovsjanikov:2012:fmaps}, given $\BPhi_k$ and $\BPsi_k$ the Laplacian bases truncated to size $k$ respectively on $\M$ and $\N$, we can project $T_{\PtoP}$ in these bases and compactly encode the functional map in a $k \times k$ matrix $\mymat{C}=\
    \BPhi_k^{\top}\ \mymat{A}_{\M}\ \PtoP\ \BPsi_k$,
which is the linear operator that transfers the coefficients of functions from $\mathcal{F}(\N)$ in the ones of corresponding functions of $ \mathcal{F}(\M)$ respectively computed with $\BPhi_k$ and $\BPsi_k$.

Finally, given a functional map $T_{\PtoP}$, or its compact representation $\mymat{C}$, we can retrieve the correspondence $\PtoP$ on which the map is built by the nearest neighbor search in the embedding space of the Laplacian eigenfunctions
as detailed in~\cite{ovsjanikov:2016:fmapsmatching}. 

\subsection{Intrinsic Delaunay triangulations (IDT)}



The IDT~\cite{rivin:1994:idt,bobenko:2007:idtlaplacian} has been introduced for generalizing the Delaunay triangulation~\cite{delaunay:1934:delaunay} to non-Euclidean metric spaces where classical algorithms~\cite{bowyer:1981:delaunay,watson:1981:delaunay} cannot be used. IDT relies on the duality with the Voronoi diagram~\cite{edelsbrunner:1994:triangulating,leibon:2000:delaunaysurf}, as three texels meeting at a point form a triangle whose edges cross the texels boundaries.
\begin{dfn}
    Let $(\M, d_{\M})$ be a 2-dimensional metric space (with distance function $d_{\M}$), and let $S = \{p_i\}_{i=1}^{s} \subset \M$ be a set of $s$ sample points. The \emph{Voronoi decomposition} (VD) of $\M$ with respect to $S$ is the collection $\{P_i\}_{i=1}^{s}$ such that
    \begin{equation}
        P_i
        =
        \left\lbrace
        q \in \M\ 
        :\ 
        d_{\M}(p_i, q) = \min_{p_j \in S}\left( d_{\M}(p_j, q) \right)
        \right\rbrace\,,
        \qquad
        \bigcup_{i=1}^{s} P_i = \M
    \end{equation}
    The points in $S$ are called \emph{generators}, and the $P_i$ are called \emph{texels} (or \emph{cells}).
\end{dfn}
The intersection between two or more texels could be non-empty. In such a case, we define those texels as \emph{adjacent}. A VD is \emph{general} if the intersection of four or more texels is empty. If a VD is general, a \emph{Voronoi vertex} is a point that belongs to three texels, and a \emph{Voronoi edge} is a curve $C$ connecting two Voronoi vertices and belonging to two texels.

It can be shown that there is a one-to-one correspondence between a Voronoi edge dividing $P_i$ and $P_j$ and a Delaunay edge connecting $p_i$ to $p_j$~\cite{dyer:2007:voronoi}. This correspondence gives rise to a necessary and sufficient condition for the IDT to be a proper triangulation (\ie, it realizes a simplicial complex).
\begin{dfn}[\cite{amenta:2000:closedball,edelsbrunner:1994:triangulating}]
    If $S$ induces a general VD, the VD satisfies the \emph{closed ball property} if:
    \begin{itemize}[noitemsep,topsep=0pt]
        \item every $P_i$ is a closed 2-ball (\ie, a topological disk without holes);
        \item every $P_i \cap P_j$ is either empty or a closed 1-ball (\ie, a topological segment);
        \item every $P_i \cap P_j \cap P_k$ is either empty or a closed 0-ball (\ie, a point).
    \end{itemize}
\end{dfn}
\begin{thm}[\cite{edelsbrunner:1994:triangulating,amenta:2000:closedball,dyer:2007:voronoi}]\label{thm:closedball}
    If $S$ induces a general VD, then the VD satisfies the closed ball property if and only if its dual IDT is a proper triangulation.
\end{thm}

\section{Method}\label{sec:method}

We introduce a new scalable functional map pipeline handling very dense meshes and yielding stable results independently of the input triangulation's quality. 

Our method acts through three main steps. At first, we compute two low-resolution meshes that preserve the original metrics of the input shapes (Section~\ref{sec:remeshing}). Then, we apply any existing pipeline, yielding a functional map between the low-resolution eigenspaces (Section~\ref{sec:matching}). Finally, we extend the eigenfunctions to the original meshes, allowing us to use them with the functional map estimated in the previous step to compute the desired correspondence between them (Section~\ref{sec:recovery}). Our technique is very general and works with arbitrary manifold triangulations, including meshes with degenerate geometry, non-orientable surfaces, and multiple connected components. The manifoldness requirement does not provide an obstacle to the applicability of our algorithm, as it can be efficiently enforced using existing techniques~\cite{fei:2013:repairnonmanifold}.

\subsection{Intrinsic Delaunay remeshing}\label{sec:remeshing}

\paragraph*{Front propagation and FPS.}

\begin{figure}[t]
    \centering
    \includegraphics[width=0.25\columnwidth]{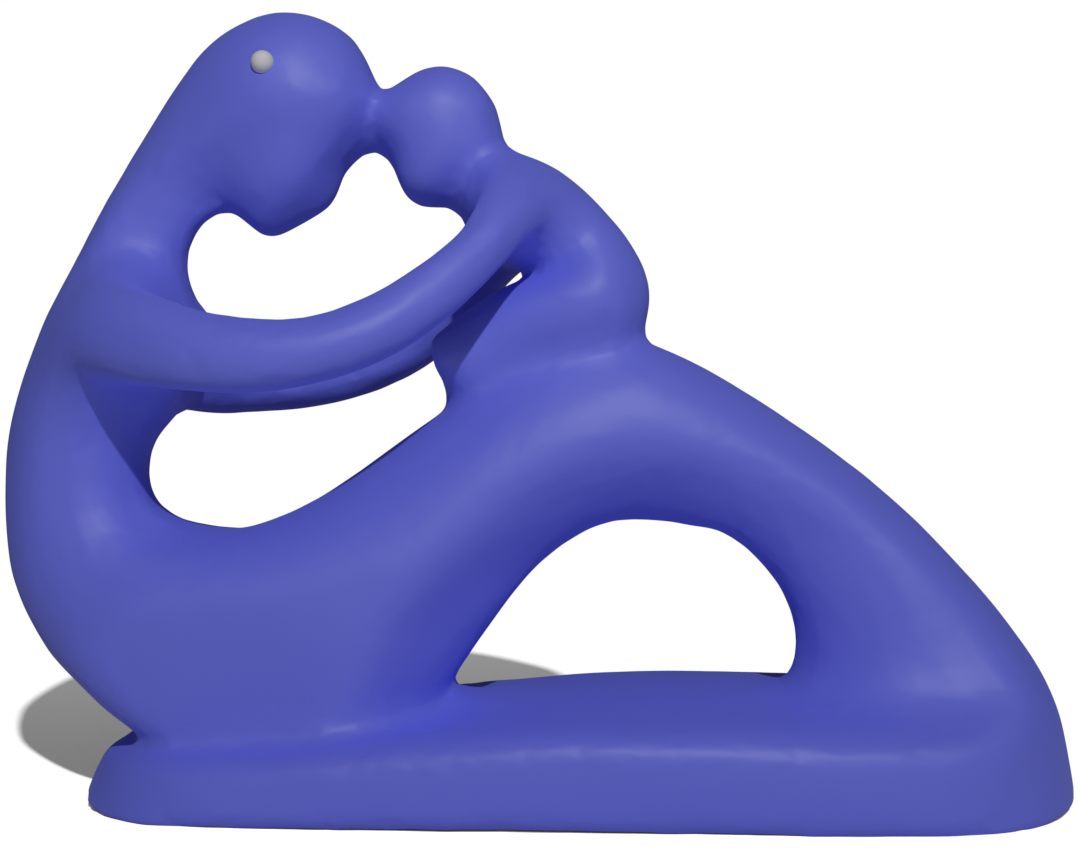}%
    \includegraphics[width=0.25\columnwidth]{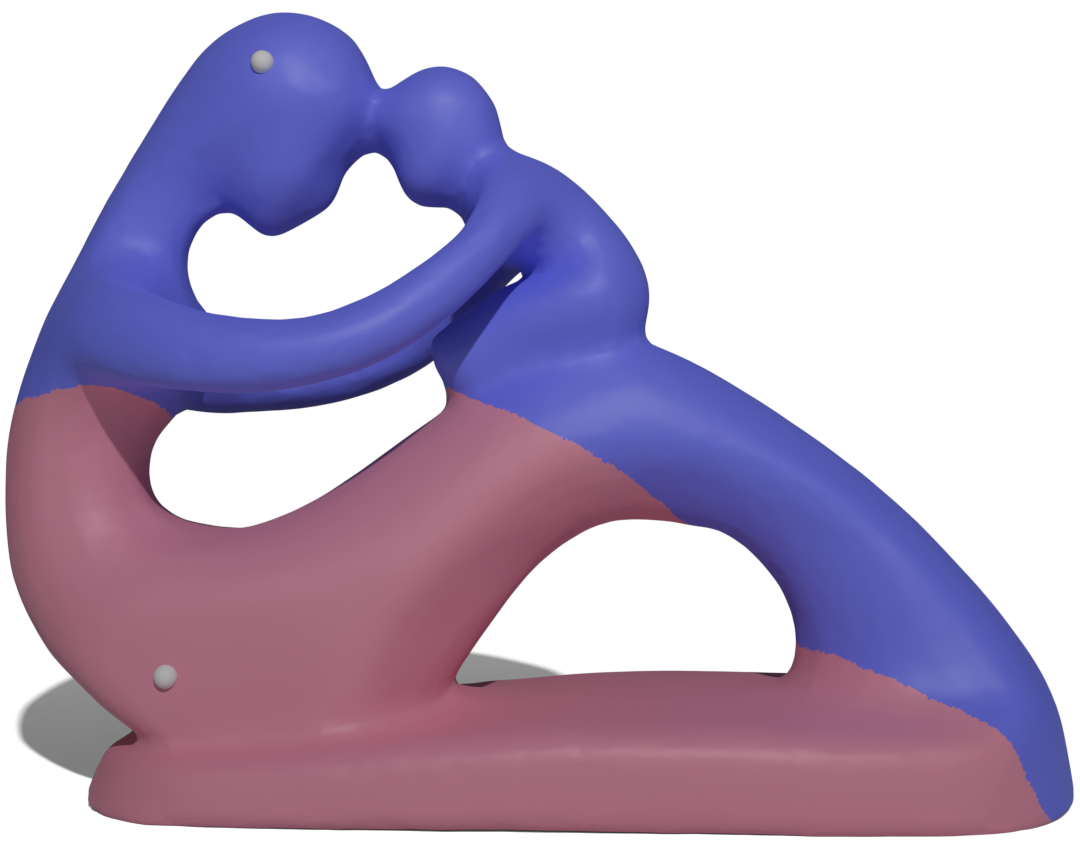}%
    \includegraphics[width=0.25\columnwidth]{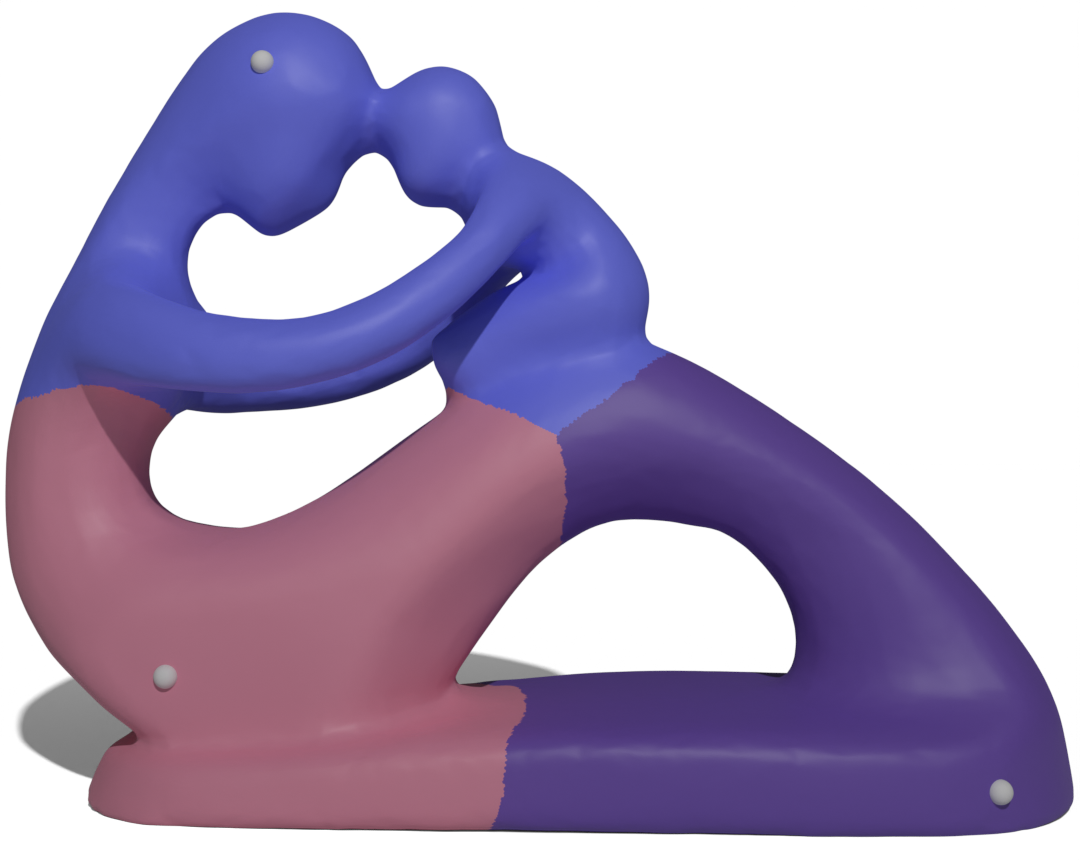}%
    \includegraphics[width=0.25\columnwidth]{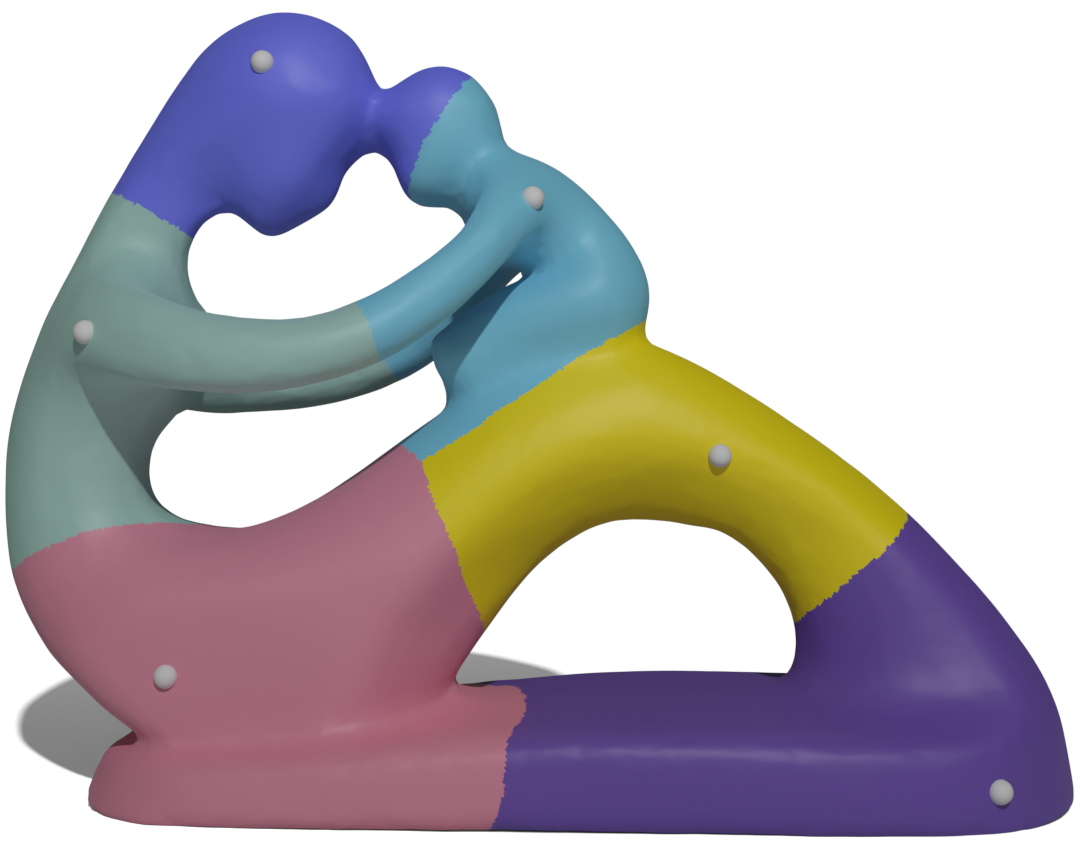}
    \caption{The iterative front propagation from samples inducing a geodesic VD.}
    \label{fig:front-propagate}
\end{figure}

Naively computing the VD induced by a set of samples can quickly become very expensive. The exact geodesic algorithm introduced by Mitchell \etal~\cite{mitchell:1987:discretegeodesics} is computationally unfeasible on large meshes. Even using faster solutions, such as the heat method~\cite{crane:2013:geodesicsisheat}, the fast marching algorithm~\cite{kimmel:1998:fastmarching}, or even the Dijkstra distance~\cite{dijkstra:1959:shortestpath}, 
searching the closest sample for each vertex 
is still a $\BigO{|V|\ s}$ operation. 
Instead, we take inspiration from the front propagation method proposed by Peyr\'{e} \etal~\cite{peyre:2006:frontpropagation}. We start with a decomposition of the mesh induced by a single sample $p_1$. This means that every vertex $v$ is part of the texel $P_1$ and has distance $D_1[v] = d_{\M}(p_1, v)$ from the sample set. We then assume to have a decomposition $P_1, \cdots, P_{k - 1}$ induced by samples $p_1, \cdots, p_{k - 1}$, and a vector $D_{k - 1}$ storing the distance of each vertex from the sample set. When adding a sample $p_k$ to the VD, we can exploit the fact that $D_{k} = \min\left( D_{k - 1}, d_{\M}(p_k, V) \right)$. We start a front from $p_k$ and expand it, updating the distances and assigning vertices to texel $P_k$ until we reach all vertices such that $D_{k - 1}[v] < d_{\M}(p_k, v)$ (see \Cref{fig:front-propagate}). By propagating the front with the fast marching or Dijkstra's algorithm, we can compute a geodesic VD with time $\BigO{|V| \log(|V|) \log(s)}$. 

To obtain the sample points, Peyr\'{e} \etal~\cite{peyre:2006:frontpropagation} propose to use a geodesic farthest point sampling, achieving a geodesic uniform sampling of the surface.
\begin{dfn}
    Let $(\M, d_{\M})$ be a 2-dimensional metric space, and let $s \in \mathbb{N}$ be an integer. A \emph{farthest point sampling (FPS)} of size $s$ with respect to $d_{\M}$ is a set $S_s \subset \M$ of $s$ 
    sample points in $\M$, incrementally built from a singleton set $S_1 = \{ p_1 \} \subset \M$ according to the following rule:
    \begin{equation}\label{eq:fpsdfn}
        S_{\ell + 1}
        =
        S_\ell
        \cup
        \left\{
        \argmax_{q \in \M}\
        \min_{p \in S_\ell}\
        d_{\M}(q, p)
        \right\}\,.
    \end{equation}
\end{dfn}
However, searching for the maximum distance in \Cref{eq:fpsdfn} at every new sample brings again the algorithm to a $\BigO{|V|\ s}$ complexity, really becoming a bottleneck for large meshes counting millions of vertices. In contrast, we propose to introduce a fixed-size binary max-heap in the front propagation algorithm, keeping track of the distances from each vertex to the sample set. During each front propagation, this heap updates the distance of each visited vertex in time $\BigO{\log(|V|)}$, adding no extra complexity to the visit. Furthermore, it can gather and set to zero the farthest vertex at each iteration in just $\BigO{\log(|V|)}$ time, totalling a time complexity of $\BigO{|V| \log(|V|) \log(s)}$ for the entire procedure. The fast front propagation algorithm is summarized in \Cref{algo:voronoi-fps}.

\begin{algorithm}[t]
\caption{Geodesic FPS and its VD.\label{algo:voronoi-fps}}
\begin{algorithmic}[1]
\Procedure{VoronoiFPS}{$\M = (V, E, T)$, $s$}
    \State $i \gets$ random index in $[1, |V|]$
    \State $S \gets \{ i \}$
    \State $D \gets d_{\M}(V, V_i)$
    \State $H \gets$ max-heap initialized with $D$
    \State $P \gets$ vector of length $|V|$ initialized to $i$
    \For{$h \gets 2$ \textbf{to} $s$}
        \State $p \gets$ \Call{FindMax}{$H$}
        \State \Call{SetKey}{$H$, $p$, $0$}
        \State $D_p \gets 0$
        \State $S \gets S \cup \{ p \}$
        \State Propagate a front from $p$.
        \State Update $D$, $P$, and $H$.
    \EndFor
\EndProcedure
\end{algorithmic}
\end{algorithm}

\paragraph*{Flat union property.}
Even if we are able to compute the VD of a farthest point sampling with high efficiency, we still cannot guarantee that the dual connectivity is a proper IDT. 
Ensuring the closed ball property can be computationally challenging. In particular, identifying the topology of the boundary between each pair of adjacent texels can become very costly when 
the number of samples is large. A possibility is to add enough samples so that the surface locally behaves like a plane. This approach, proposed by Leibon \etal~\cite{leibon:2000:delaunaysurf} and adopted by Peyr\'{e} \etal~\cite{peyre:2006:frontpropagation}, makes it easy to lose control over the number of samples, and anyway the requirements are not easy to enforce. 
For guaranteeing a proper IDT, while still avoiding these issues, we introduce a novel property for a VD. The proofs of our claims are provided in \Cref{sec:thmproof,sec:prpproof}.
\begin{dfn}[Flat Union Property (FUP)]
    Let $(\M, d_{\M})$ be a 2-dimensional metric space. Let then $S = \{p_i\}_{i=1}^{s} \subset \M$ be a set of $s$ sample points in $\M$, inducing a general VD $\{P_i\}_{i=1}^{s}$. The VD induced by $S$ satisfies the \emph{FUP} if:
    \begin{itemize}
        \item every $P_i$ is a closed 2-ball;
        \item if $P_i \cap P_j$ is not empty, then $P_i \cup P_j$ is a closed 2-ball;
        \item if $P_i \cap P_j \cap P_k$ is not empty, then $P_i \cup P_j \cup P_k$ is a closed 2-ball.
    \end{itemize}
\end{dfn}
\begin{thm}\label{thm:flatunion}
    Let $(\M, d_{\M})$ be a 2-dimensional metric space. If a general VD $\{P_i\}_{i=1}^{s}$ of $\M$ satisfies the \emph{FUP}, it also satisfies the closed ball property.
\end{thm}

The FUP is much easier to verify than the closed ball property, as we only have to ensure that regions are closed 2-balls. Indeed, the following property can be exploited in this regard. 
\begin{prp}\label{prp:2ball}
    Let $\M = (V, E, F)$ be a manifold polygonal mesh, then $\M$ is a closed 2-ball if and only if its Euler characteristic $\chi = |V| - |E| + |F|$ is 1.
\end{prp}


\paragraph*{Dual mesh representation.}
When we compute a VD of a triangle mesh, we want to partition the vertices into disjoint sets. However, Voronoi texels are defined as regions over the surface, meaning that they should be submanifolds of $\M$. By only considering vertices, we likely end up having non-manifold geometries or non well-defined boundaries.

\begin{figure}[t]
    \centering
    \includegraphics[width=0.25\columnwidth]{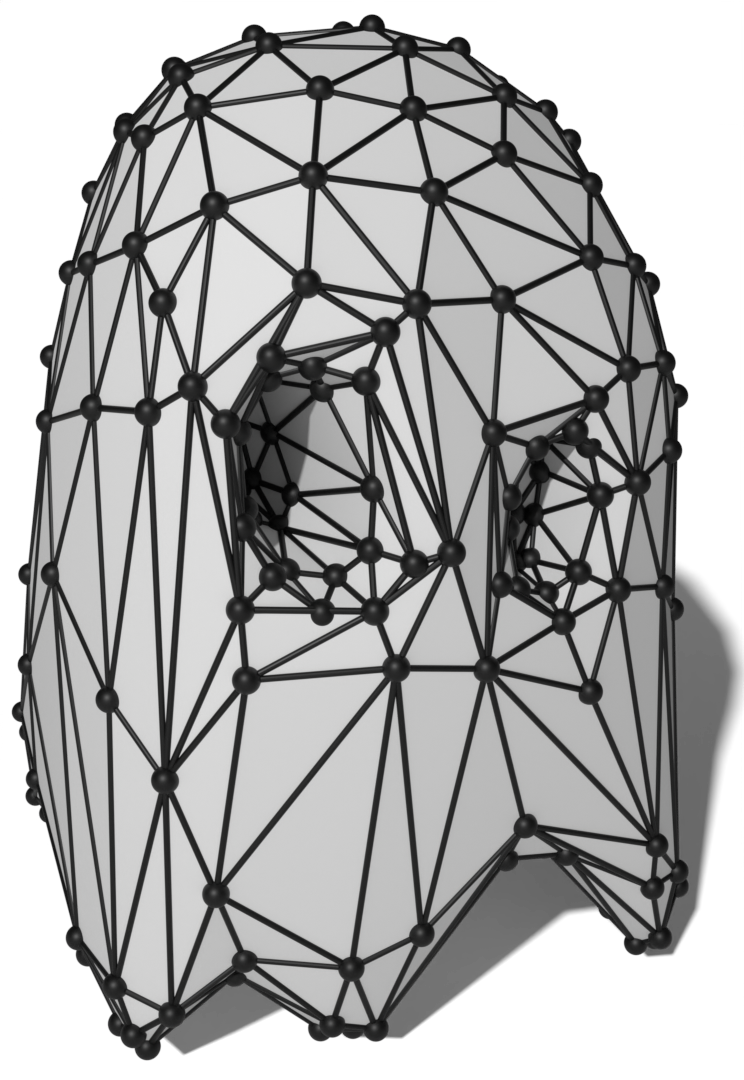}%
    \includegraphics[width=0.25\columnwidth]{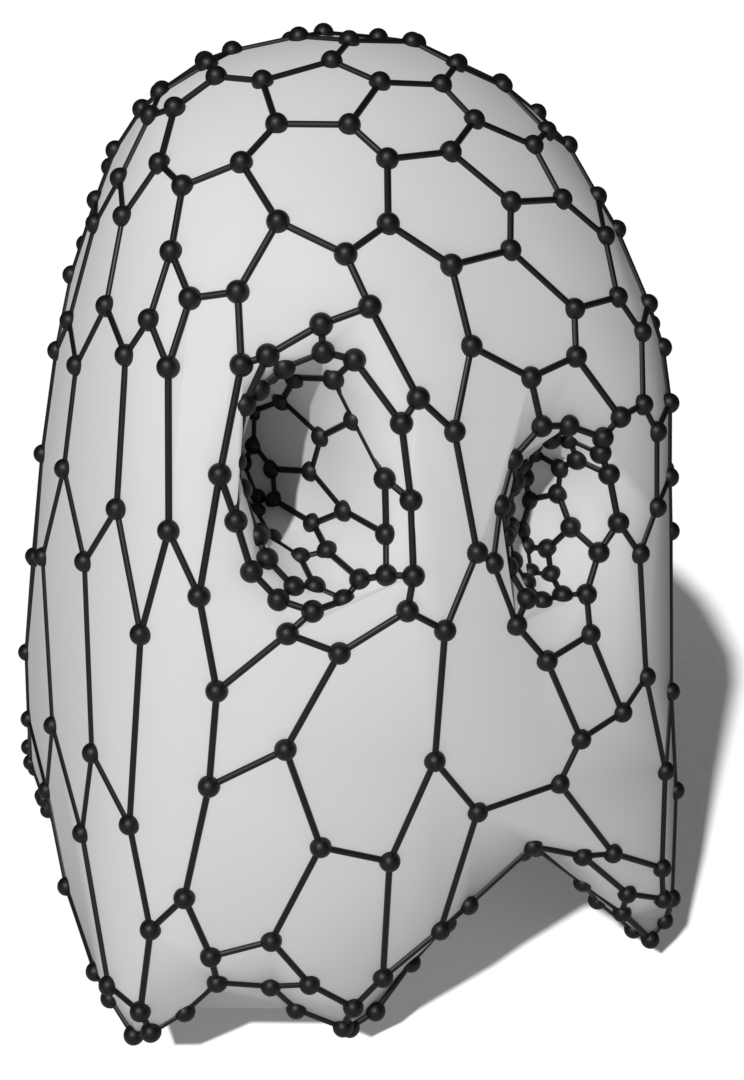}%
    \includegraphics[width=0.25\columnwidth]{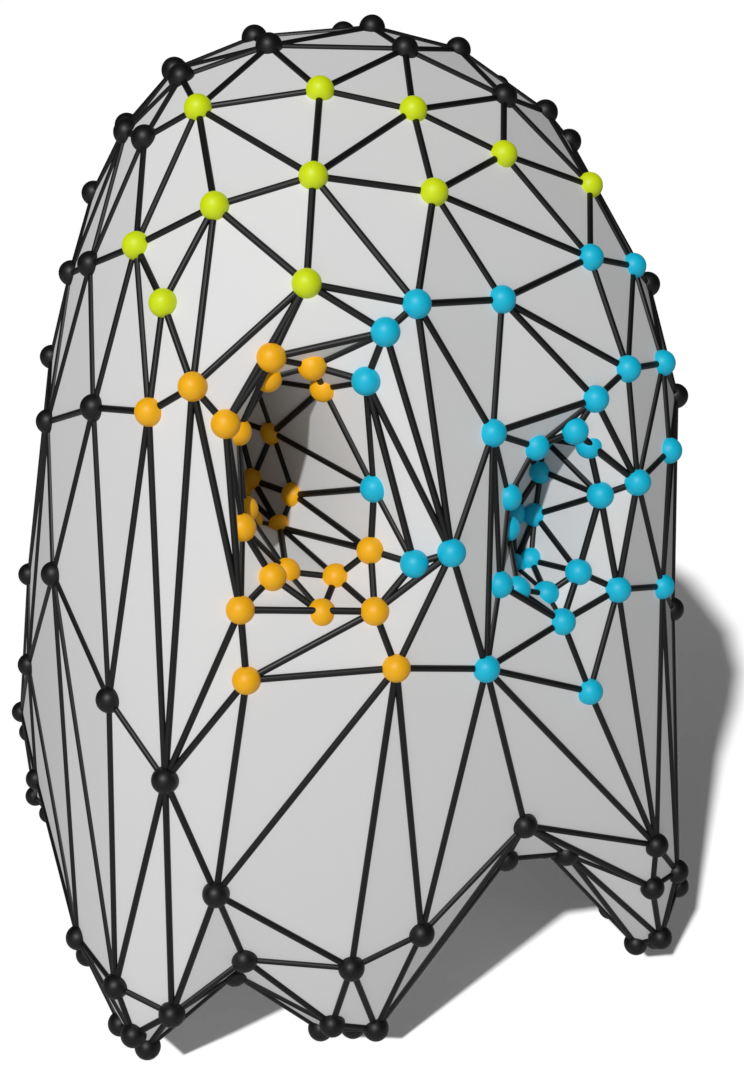}%
    \includegraphics[width=0.25\columnwidth]{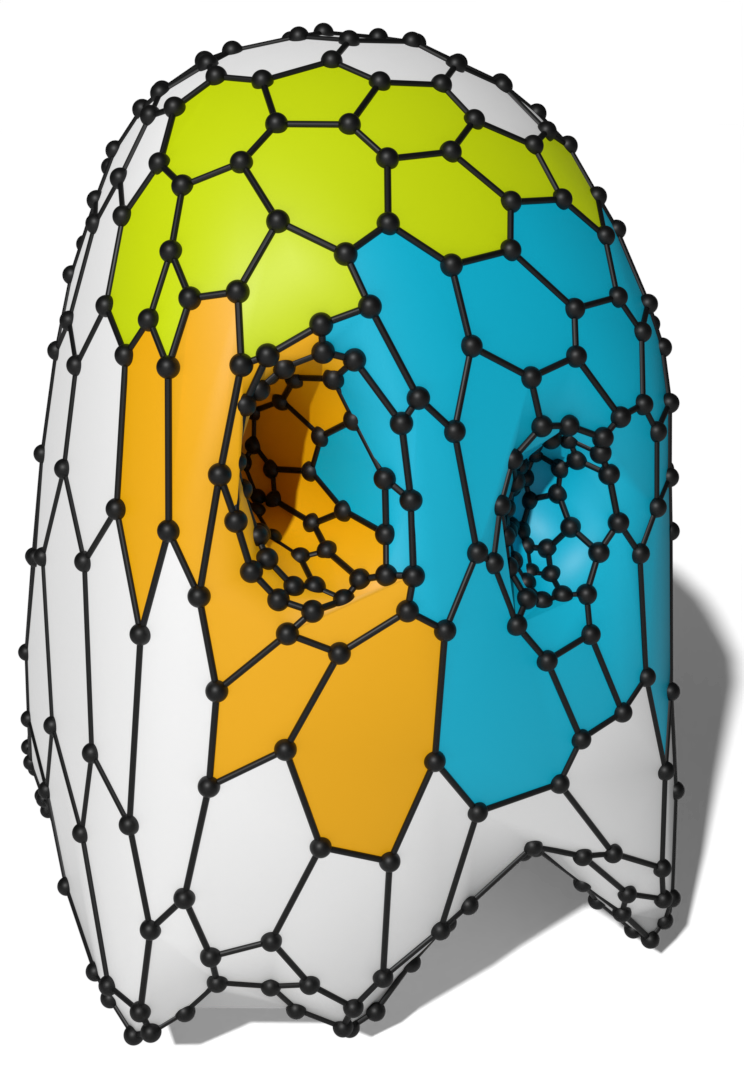}
    \vspace{-0.5em}
    \caption{Top: the primal and dual connectivities of a triangular mesh. Bottom: regions on the primal mesh are represented as sets of vertices, and the corresponding regions on the dual mesh are composed by faces.}
    \label{fig:primal-dual}
\end{figure}

We consider the polygonal mesh $\Tilde{\M} = (\Tilde{V}, \Tilde{E}, \Tilde{T})$ to be the dual mesh of $\M = (V, E, T)$, built by placing a vertex at each face of $\M$ and forming a polygonal face for each triangle fan around a vertex of $\M$. As shown in the example from \Cref{fig:primal-dual}, dual meshes allow us to define Voronoi texels by means of connected dual faces, making them actual submanifolds of the original surface. Furthermore, we can easily define boundaries between texels as paths made of dual edges.

This construction has the positive side effect to ensure that every VD is general. Indeed, the only place where three or more texels can meet is a dual vertex. Since texels are made of dual faces (\ie, primal vertices), and the primal mesh is triangular, there is no dual vertex where more than three faces can meet.

Given the correspondence between primal and dual geometric elements, we can verify if a region is a closed 2-ball with \Cref{prp:2ball} without explicitly constructing the dual mesh. Exploiting efficient data structures, like hash-maps and hash-sets, we can verify the topology of each texel (or union of two or more texels) with a single pass over vertices, edges and triangles. After the initial FPS, we further add samples to enforce the flat union property: if two or more adjacent texels do not form a closed 2-ball, we break the connection adding samples at the Voronoi vertex or along the Voronoi edge; if a texel is not a closed 2-ball, we add a sample along its boundary to reduce its coverage.



\subsection{Low-resolution matching}\label{sec:matching}

Let $\M$ and $\N$ be two triangular meshes, and let $\DiscM$ and $\DiscN$ be their low-resolution counterparts computed with the algorithm presented above (or any other alternative algorithm). Furthermore, let $\DBPhi$ be a basis for the functional space over $\DiscM$, and $\DBPsi$ a basis for the functional space over $\DiscN$. For example these bases can be the truncated subset of the eigenvectors of the LBO as proposed in~\cite{ovsjanikov:2012:fmaps}. We assume to have some pipeline yielding a functional map $\mymat{C}$, such that
\begin{equation}\label{eq:funmap-lowres}
    \DBPsi\ \mymat{C}\
    =\
    \LRPtoP\ \DBPhi\,,
\end{equation}
where $\LRPtoP : \DiscM \to \DiscN$ is a correspondence between the vertices of $\DiscM$ and the vertices of $\DiscN$.
The adopted functional maps solution can be any of the available alternatives, ranging from the original one~\cite{ovsjanikov:2012:fmaps} to the most recent~\cite{donati:2022:complexfmaps}. In our experiments, we consider two of the widely adopted solutions: constrained optimization with product preservation~\cite{nogneng:2017:descriptorpreserve} and the iterative procedure ZoomOut~\cite{melzi:2019:zoomout}. The first is representative of the possible optimization strategies, while the second is probably the most efficient refinement technique for functional maps estimation. Both these methods have been efficiently and effectively applied on meshes with a limited number of vertices (\ie, up to 10 thousands vertices), which is exactly the setting we are in after the proposed remeshing step.

\subsection{Extending the correspondence}\label{sec:recovery}

\begin{figure}[t]
    \centering
    \includegraphics[width=\columnwidth]{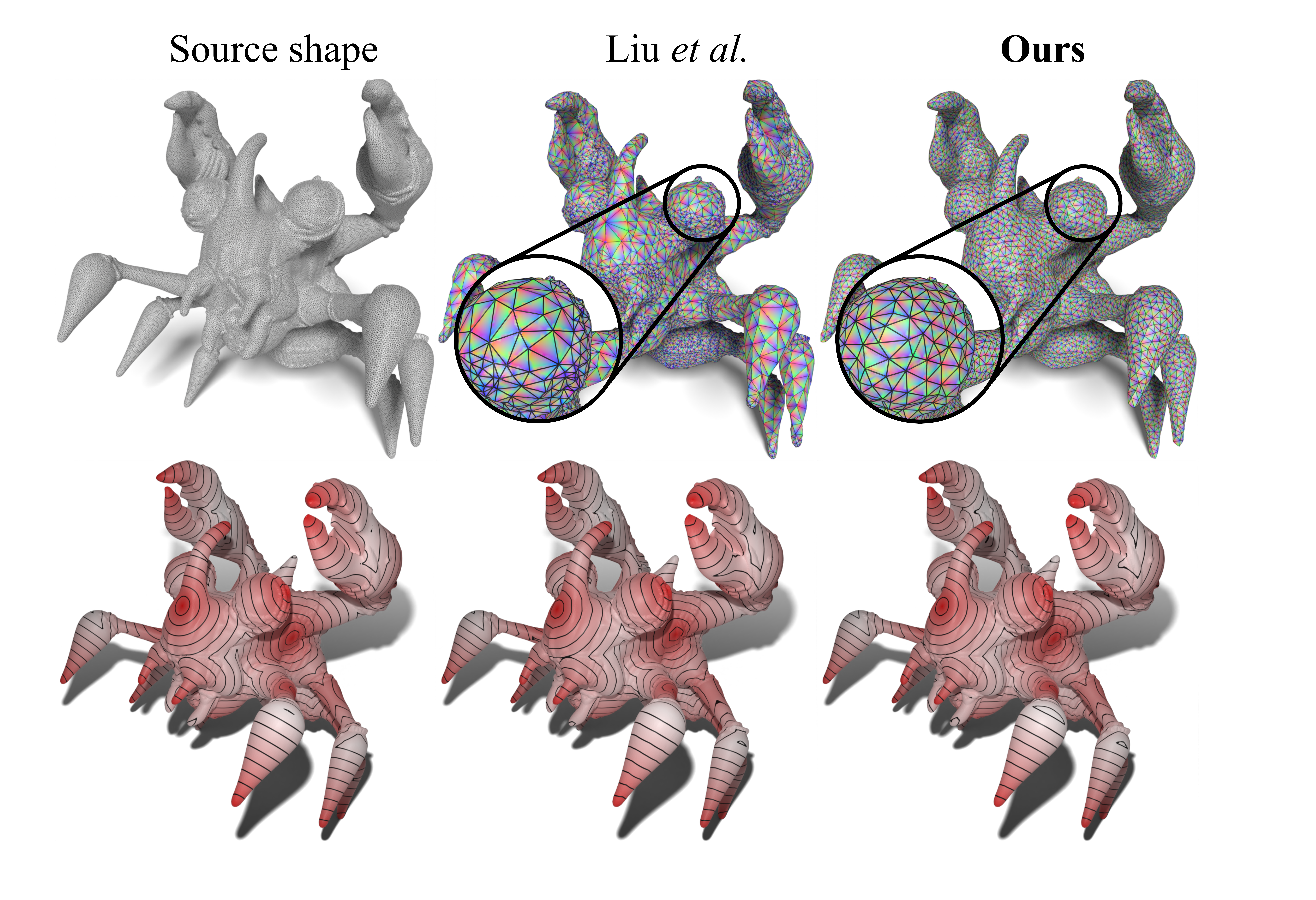}
    \caption{Top: remeshing and mapping of the original vertices with the solution from Liu \etal~\cite{liu:2023:intrinsicsimplification} and our pipeline. Bottom: comparison of ground truth geodesic distances from multiple vertices and extended with both methods.}
    \label{fig:hi2low-mapping}
\end{figure}

To extend the basis from the remeshing to the original surface, we need a mapping that transports scalar fields from the low-resolution mesh $\DiscM$ to the high-resolution shape $\M$. 
A possible solution would be to map each triangle of $\DiscM$ to a geodesic triangle in $\M$, and then computing the geodesic barycentric coordinates of each vertex inside the triangle~\cite{liu:2023:intrinsicsimplification}. Unfortunately, this approach is very costly, as it requires to compute exact geodesic paths onto $\M$~\cite{rustamov:2010:barycentric}.
Instead, we approximate this mapping by projecting each vertex of $\M$ onto the closest surface point of $\DiscM$. Despite this approach not being accurate in general, every triangle in $\DiscM$ corresponds to a geodesic triangle (homeomorphic to a disk) on $\M$ with the same vertices. Furthermore, starting from a farthest point sampling gives evenly spaced samples, mitigating the geometric complexity of geodesic triangles.

A surface point of $\DiscM$ is either a vertex, or a weighted average of two (if on an edge) or three vertices (if on a triangle). We build a linear map $\mymat{U}_{\M} \in \Reals^{m \times s}$ representing the projection of each vertex $ v_i \in \M$ as a linear combination of at most three vertices in $\DiscM$ (\ie, the vertices encoding the closest surface point).
\Cref{fig:hi2low-mapping} shows a comparison between our pipeline and the simplification method proposed by Liu \etal~\cite{liu:2023:intrinsicsimplification}. The top row shows the different geometries produced by the two methods in reducing the mesh size by 90\% and how our mapping compares to the approximate geodesic barycentric coordinates. In the bottom row we compare the two methods in approximating a multiple source geodesic distance field, showing that they do not present appreciable differences. 
More discussion is provided in \Cref{sec:mapping}.

Scalar functions can be evaluated at any surface point by interpolating the values at the vertices. Since every vertex of $\M$ is associated to a point on the surface of $\DiscM$, we can use the linear map $\mymat{U}_{\M}$ for extending scalar functions from the $s$ vertices of $\DiscM$ to the $m$ vertices of $\M$. 

We extend the bases $\DBPhi$ and $\DBPsi$ to the full-resolution meshes as $\BPhi = \mymat{U}_{\M}\DBPhi$ and $\BPsi = \mymat{U}_{\N}\DBPsi$. Then, in a similar fashion as done in Equation \eqref{eq:funmap-lowres}, we search for a correspondence $\PtoP : \M \to \N$ that maps vertices of $\M$ to vertices of $\N$ as
\begin{equation}\label{eq:funmap-extension}
    \mymat{U}_{\N}\ \DBPsi\ \mymat{C}\
    =\
    \BPsi\ \mymat{C}\
    =\
    \PtoP\ \BPhi\
    =\
    \PtoP\ \mymat{U}_{\M}\ \DBPhi\,,
\end{equation}
where, for the sake of clarity, we explicitly write the equation in terms of the bases $\DBPhi$ and $\DBPsi$.
Given the extended bases $\BPhi$ and $\BPsi$, and the same functional map $\mymat{C}$ estimated to solve Equation \eqref{eq:funmap-lowres}, we obtain the correspondence $\PtoP$ via a nearest neighbor search in the spectral space. 

We notice that nearest neighbor algorithms become very slow on large meshes, effectively becoming a possible bottleneck for the pipeline. Our experiments presented in \Cref{fig:shrec19-remesh-times} show that, to the scope of this paper, this solution provides satisfying performance.
However, we acknowledge that establishing a relationship between \eqref{eq:funmap-lowres} and \eqref{eq:funmap-extension} could lead to a more efficient expression of $\PtoP$ in terms of $\LRPtoP$, $\mymat{U}_{\M}$, and $\mymat{U}_{\N}$. This avenue warrants future exploration.


\section{Results}\label{sec:results}


Our goal is to evaluate not only the matching accuracy of our method but also its time performance. For this task, we test it on datasets for dense correspondences, such as the SHREC19 dataset~\cite{melzi:2019:shrec19} and the TOSCA dataset~\cite{bronstein:2008:tosca}. We also introduce a novel dataset, BadTOSCA, obtained by randomly altering the vertex positions of the TOSCA meshes, to ensure that our method is stable even under circumstances where strong isometry cannot be assumed\footnote{Details on the dataset generation are provided in \Cref{sec:badtosca}.}.

\begin{figure*}[t]
    \centering
    \includegraphics[width=\columnwidth]{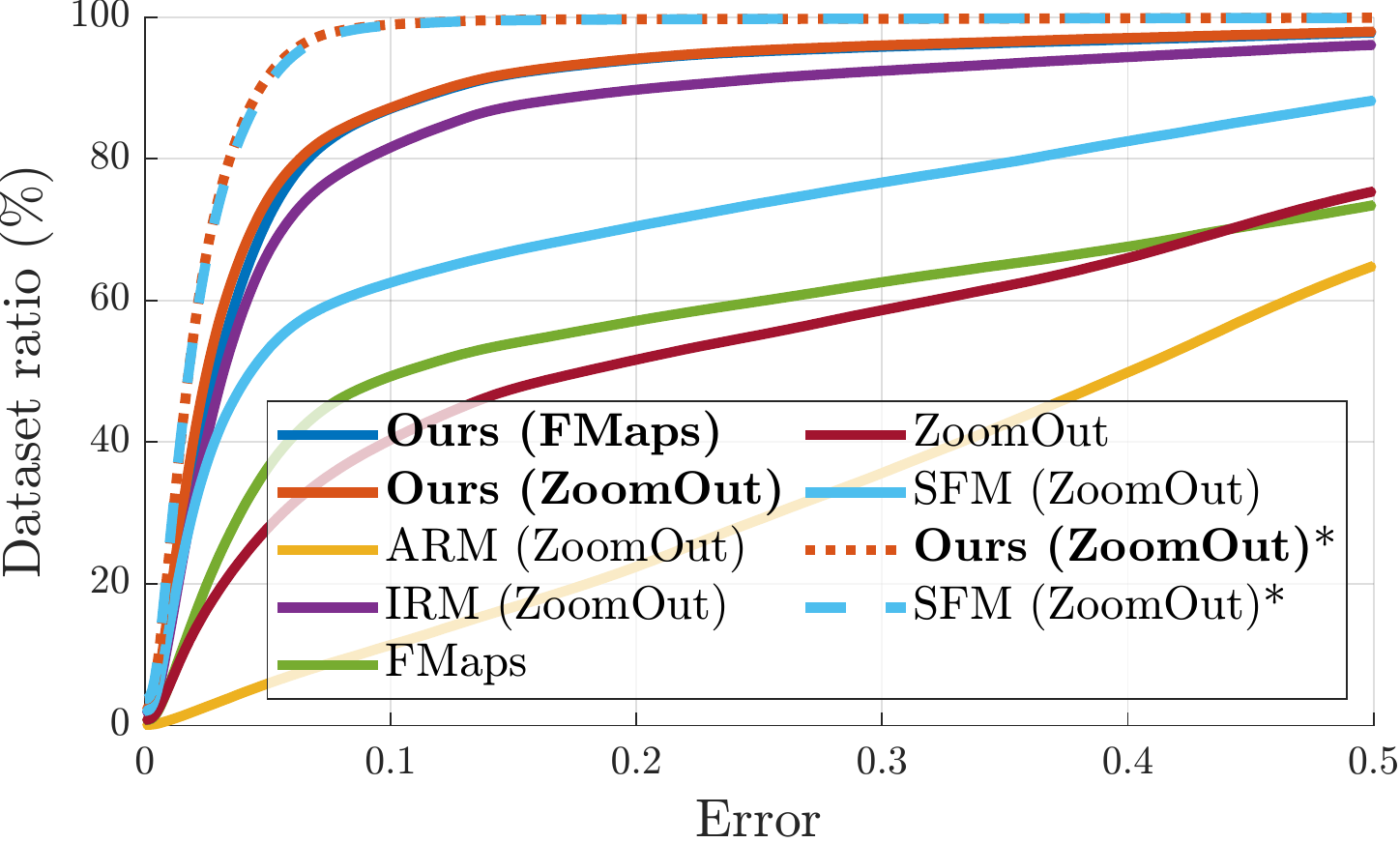}%
    \includegraphics[width=\columnwidth]{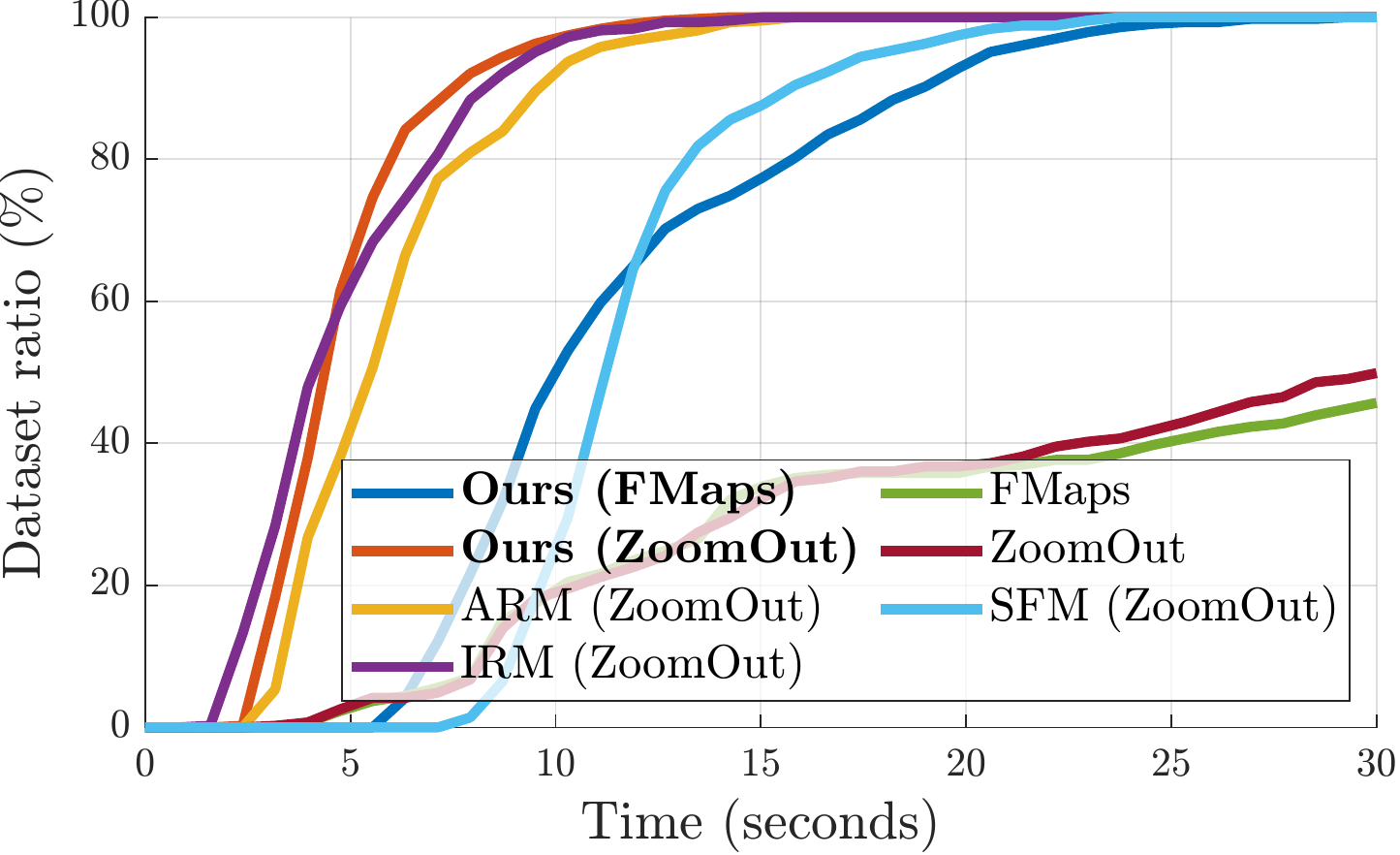}
    \caption{Left: Accuracy curves on the SHREC19 challenge pairs for the tested methods. We also consider ideal approaches where the initial functional map for ZoomOut is given ground truth dashed lines). Right: Cumulative curves of the execution time for evaluated methods. For all methods, we also consider the time required to run the remeshing or resampling step(s).}
    \label{fig:shrec19-geoderror}
\end{figure*}

Our pipeline relies on an efficient remeshing algorithm that can quickly and drastically reduce the size of a mesh by orders of magnitude while still preserving the underlying geometry and metric. In principle, any remeshing algorithm could be adopted in our pipeline, as the map for extending scalar fields discussed in \Cref{sec:recovery} only requires having two input meshes. We compare our solution against well-established isotropic and anisotropic remeshing algorithms (respectively, IRM~\cite{hoppe:1993:meshopt,botsch:2006:geometric} and ARM~\cite{nivoliers:2015:anisotropic}), as well as the scalable functional maps (SFM) approach proposed by Magnet \etal~\cite{magnet:2023:scalablefmaps}, which produces a subsampling of the vertices in place of a low-resolution mesh.

We implemented our remeshing algorithm in C++, using Eigen~\cite{eigenweb} and libigl~\cite{libigl}, and the matching pipeline in MATLAB.

\subsection{Quality of matching}\label{sec:results:quality}

To evaluate our shape matching pipeline, we tested our method on the SHREC19 dataset~\cite{melzi:2019:shrec19}. 
We can use different shape correspondence techniques to test the entire pipeline. In particular, we compute the functional map on the low-resolution representation using both the widely adopted approach with products preservation (FMaps)~\cite{nogneng:2017:descriptorpreserve} and ZoomOut~\cite{melzi:2019:zoomout}. We compare our method with the same approaches on the full-resolution meshes, as well as against SFM. 
For a fair comparison, we remesh all the shapes to 3k vertices with all methods, as Magnet \etal~\cite{magnet:2023:scalablefmaps} state that this produces the best performance with SFM. To further ease the alignment of the Laplacian spectra, we also post-process the remeshed shapes (produced with IRM, ARM, and our algorithm) to remove small disconnected components made of few triangles before computing the extension map discussed in \Cref{sec:recovery}.
This post-processing step cannot be performed for SFM, due to the nature of its sampling strategy. The benefits of this post-processing will be discussed in \Cref{sec:results:benefits}.

\begin{table}[t]
    \centering
{
\begin{tabular}{ccc}
    Method  & AGE ($\cdot 10^{-2}$) $\downarrow$  & AUC $\uparrow$    \\
    \hline
    \TableFirst{\textit{\textbf{Ours (ZoomOut)*}}}    
    & \TableFirst{\textbf{\textit{2.40}}} 
    & \TableFirst{\textbf{\textit{95.12\%}}}    \\
    
    \TableSecond{\textit{SFM*}}                                   
    & \TableSecond{\textit{2.46}}               
    & \TableSecond{\textit{95.22\%}}              \\
    
    \hline
    
    \TableSecond{Ours (FMaps)}                              
    & \TableSecond{6.23}                        
    & \TableSecond{88.21\%}                       \\
    
    \TableFirst{\textbf{Ours (ZoomOut)}}                   
    & \TableFirst{\textbf{5.87}}               
    & \TableFirst{\textbf{88.82\%}}              \\
    
    SFM                                             
    & 16.84                       
    & 70.19\%                       \\
    
    ARM (ZoomOut)                                   
    & 41.59                       
    & 30.34\%                       \\
    
    IRM (ZoomOut)                                   
    & 8.30                        
    & 84.62\%                       \\
    
    FMaps                                           
    & 28.31                       
    & 56.15\%                       \\
    
    ZoomOut                                         
    & 29.36                       
    & 52.02\%            
    
\end{tabular}%
}
    \caption{Average geodesic error and area under the accuracy curve for each tested method on the SHREC19 challenge pairs. The top rows (denoted by $*$) show the performance of our algorithm and SFM under the assumption of having a ground truth functional map for initializing ZoomOut.}
    \label{tab:shrec19-auc-age}
\end{table}

We summarize the results of our experiment in \Cref{fig:shrec19-geoderror}, where we show the accuracy curves for all methods on the SHREC19 connectivity track benchmark, following the paradigm proposed by Kim \etal~\cite{kim:2011:bim}. To prove the time efficiency of our pipeline, we also measure the time taken by every method on each pair of shapes in the benchmark (including the remeshing step), and we show the cumulative curves of the execution times over the entire dataset in a similar way with respect to the accuracy curves.

We see that all the low-resolution approaches provide better time and quality performance than the full-resolution methods, as the simpler geometry makes it easier and faster to align the Laplacian eigenfunctions. ARM is the only exception, since the anisotropic remeshing produces many skew triangles and a singular Laplacian matrix for all the shapes.
Furthermore, our technique (with both the FMaps and ZoomOut backend) outperforms SFM and achieves better results than using the IRM remeshing strategy. This difference can be better appreciated in \Cref{tab:shrec19-auc-age}, where we report the average geodesic error (normalized with respect to the shape diameter) and the area under the accuracy curves. In both \Cref{fig:shrec19-geoderror} and \Cref{tab:shrec19-auc-age} we have also considered the idealized scenario where our method with ZoomOut as backend and SFM are initialized with a ground truth functional map. In this case, the two approaches obtain comparable results.

\begin{figure*}[t]
    \centering
    \includegraphics[width=\columnwidth]{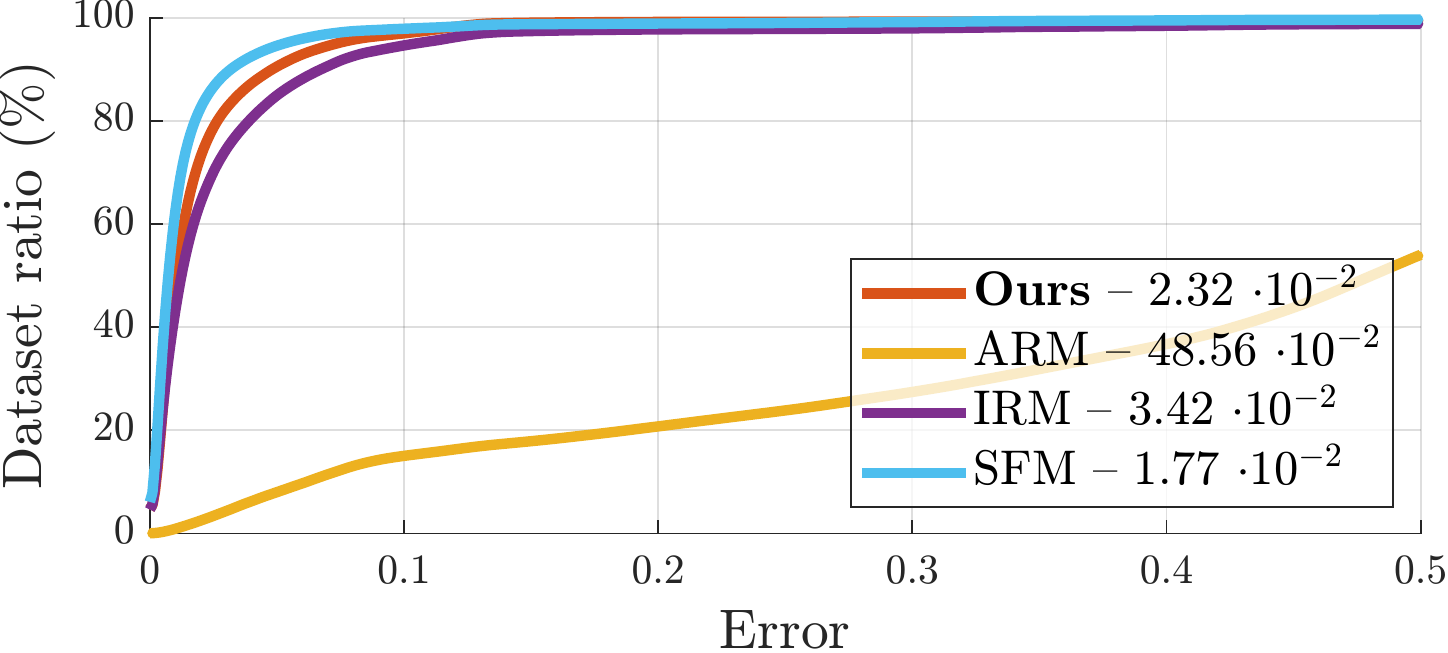}%
    \includegraphics[width=\columnwidth]{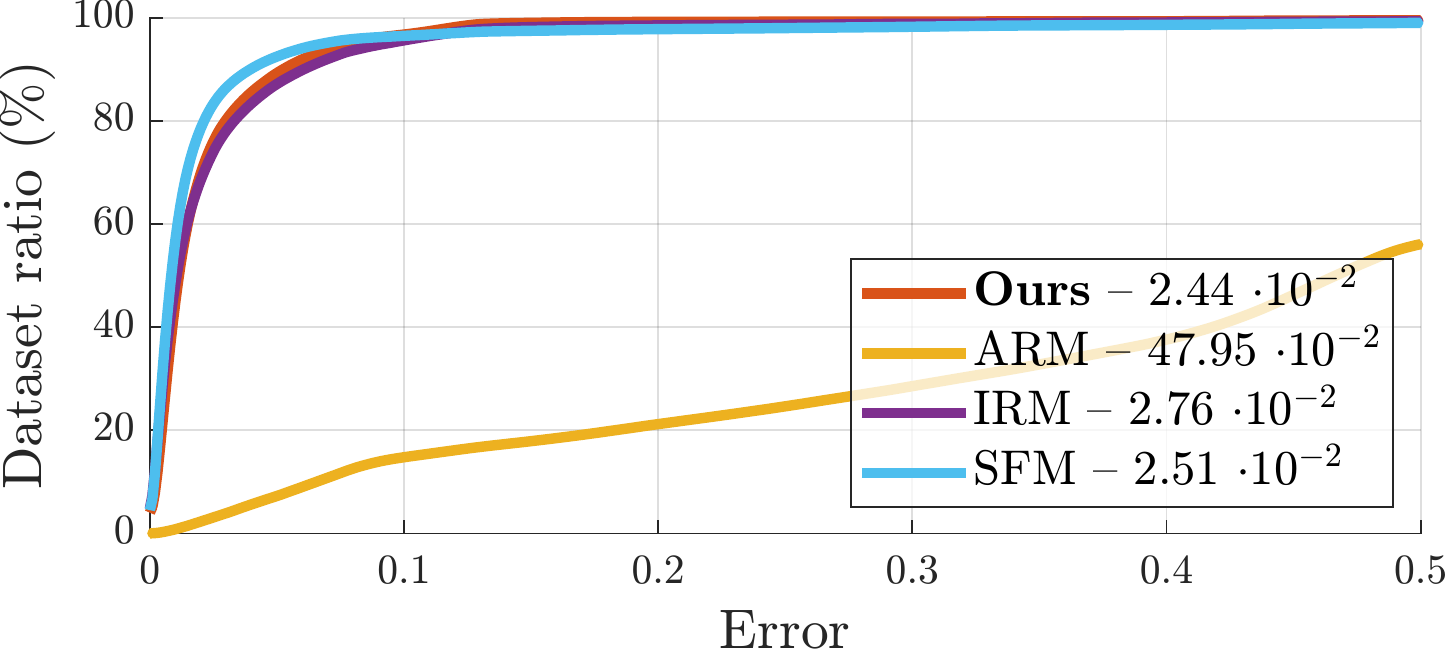}
    \caption{Accuracy curves and average geodesic error on the TOSCA dataset (left) and the BadTOSCA dataset (right) for the tested methods.}
    \label{fig:tosca-geoderr}
\end{figure*}

We also compare our method with the other approaches on the TOSCA and BadTOSCA datasets, summarizing the results in \Cref{fig:tosca-geoderr}. In the figure, we show the accuracy curves for the tested methods on both datasets, also providing the average geodesic error normalized by the shape diameter. We see that under the favorable and unrealistic conditions offered by TOSCA, where shapes of the same class are almost perfectly isometric, SFM achieves the best results. However, when we consider the slightly altered geometry of BadTOSCA, SFM suffers a performance drop of about 50\%, while our method shows more stability.

\begin{figure}[t]
    \centering
    \includegraphics[width=\columnwidth]{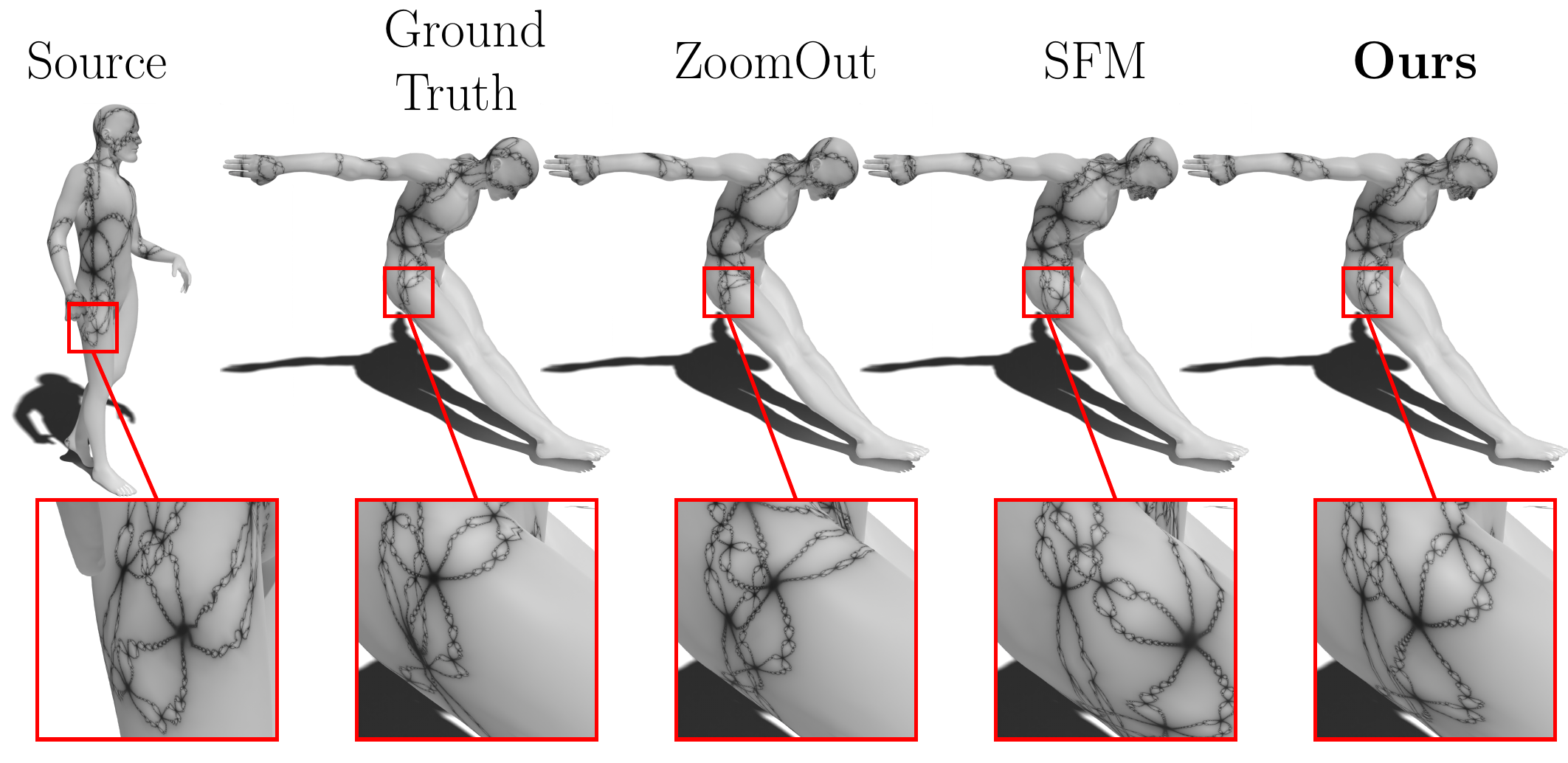}
    \caption{Coordinate transfer between non-isometric shapes. The coordinates are used to generate highly complex patterns, whose sensitivity to the input enhances the transfer errors.}
    \label{fig:shrec19-transfer-newton}
\end{figure}

A significant benefit of the functional map approach is its independence from the difference in resolution between the source and target shape. Transferring a function with functional maps yields a much smoother and neat result than directly a point-wise correspondence. 
In this regard, we test our method in transferring coordinate functions between isometric and non-isometric shapes with substantial differences in resolution and triangulation. In the example from \Cref{fig:shrec19-transfer-newton} we generated a complex procedural fractal pattern as a function of the coordinates~\cite{maggioli:2022:newton}. This pattern is very sensitive to the input, so even slight errors are highly enhanced. We see that SFM cannot precisely map the coordinates of the source mesh, shifting around and distorting the details of the pattern.

\begin{figure}[t]
    \centering
    \includegraphics[width=\columnwidth]{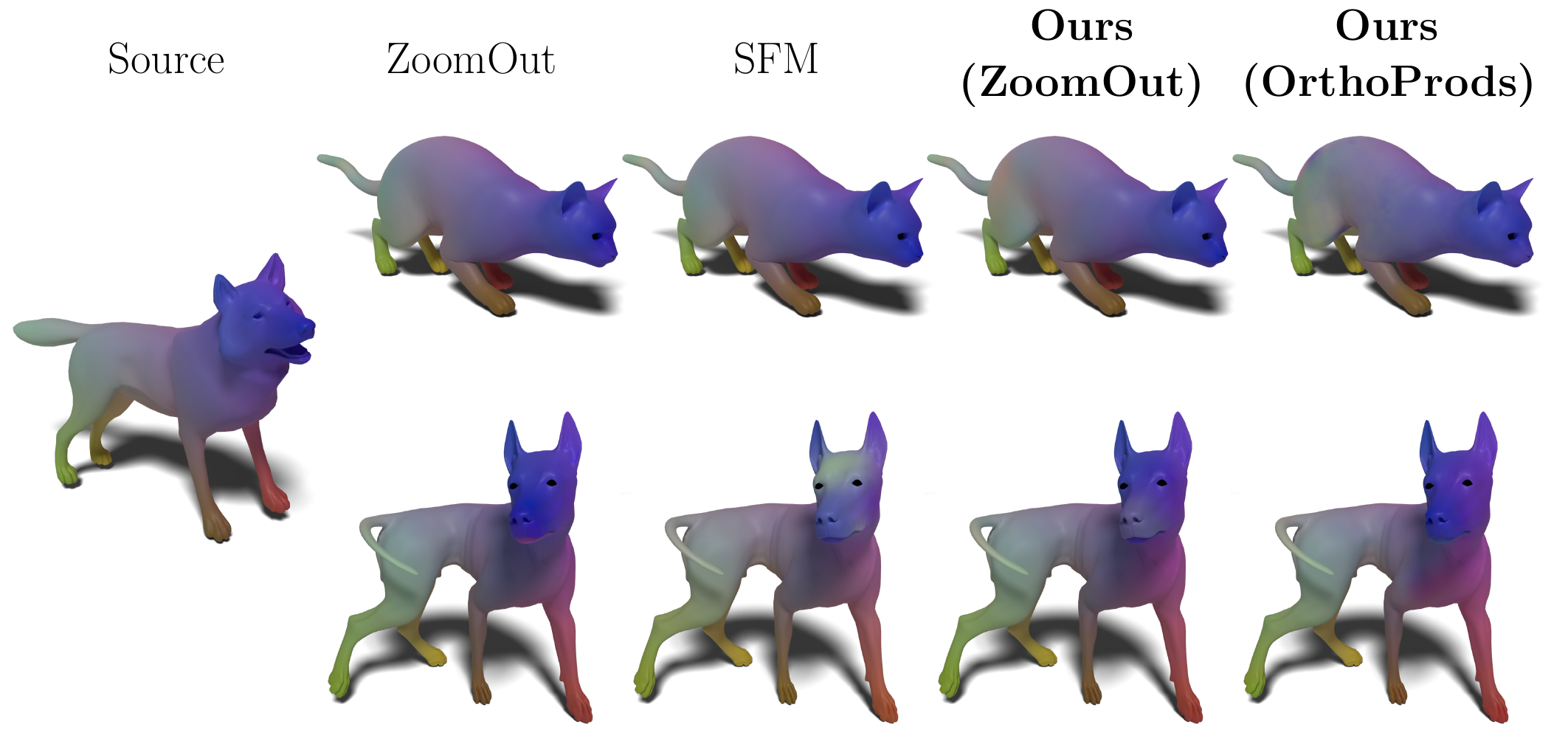}
    \caption{Inter-class coordinate transfer from the wolf model to a cat and a dog. Our pipeline is not constrained by the standard Laplacian basis and can be used with other approaches as well.}
    \label{fig:tosca-interclass}
\end{figure}

Furthermore, the example in \Cref{fig:tosca-interclass} shows that we can also deal with inter-class function transfer with our pipeline, outperforming SFM in a visual comparison. In the example, we also showcase that we should not necessarily rely on the standard Laplacian basis as a backend. Instead of using ZoomOut to transfer functions between the low-resolution meshes, we rely on the orthogonalized eigenproducts basis introduced by Maggioli \etal~\cite{maggioli:2021:orthoprods}, extending it with the same technique described in \Cref{sec:matching}. 

\begin{figure}[t]
    \centering
    \includegraphics[width=\columnwidth]{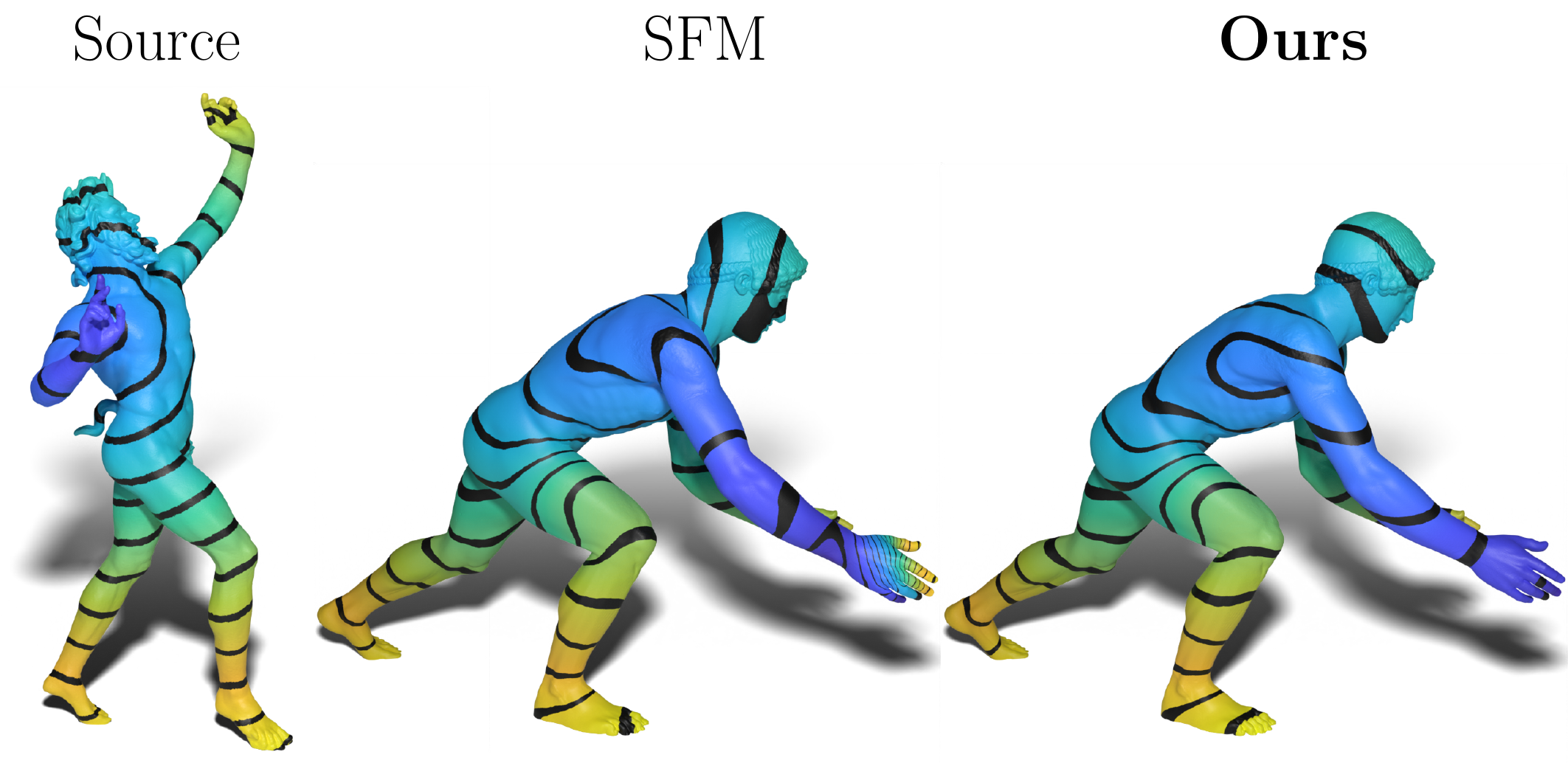}
    \caption{Comparison of function transfer between very high-resolution models with SFM and our approach.}
    \label{fig:statues-compare}
\end{figure}

Finally, to ensure that our method can scale to high-resolution meshes, we tested it using two very high-resolution 3D scans of real statues: a dancing faun counting $\sim$750k vertices and a warrior from the temple of Aphaea counting $\sim$3.5M vertices (see Figures \ref{fig:teaser} and  \ref{fig:statues-compare}). Since SFM can also deal with dense shapes, we compare its results with the ones of our method. \Cref{fig:statues-compare} shows an example of the transfer of a geodesic distance function. While our method can faithfully transfer the function, SFM produces incoherent distances (\eg, wrong isolines on the head and near the shoulder) and evident artefacts (\eg, the wrong distance on the hand).

\subsection{Scalable performance}\label{sec:results:performance}




\begin{figure}[t]
    \centering
    \includegraphics[width=\columnwidth]{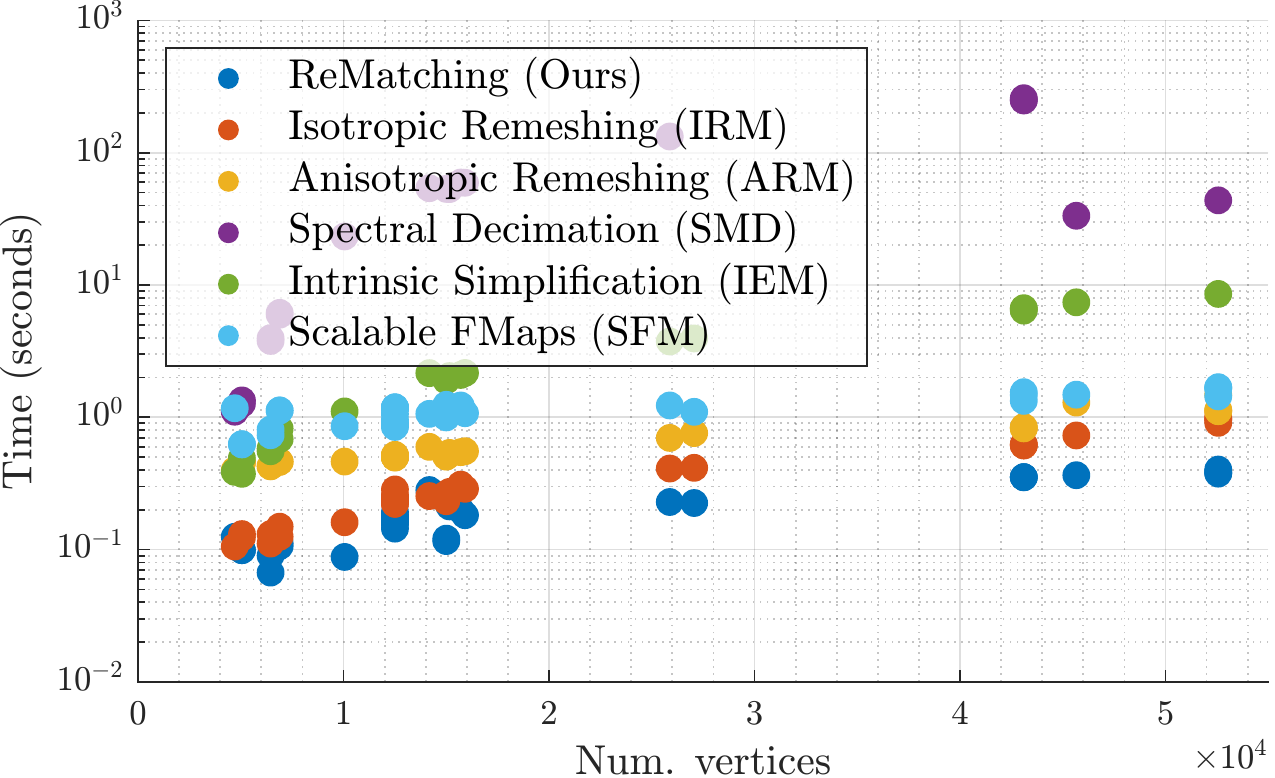}
    \caption{Execution time (logarithmic scale on the $y$-axis) of different remeshing algorithms plotted against the number of vertices of the input shape ($x$-axis). For SFM we show the time taken for generating the sampling.
    }
    \label{fig:shrec19-remesh-times}
\end{figure}

In \Cref{sec:results:quality} we saw that general-purpose remeshing solutions are not an optimal choice for the matching task. However, Lescoat \etal~\cite{lescoat:2020:spectraldecimation} provided a specialized mesh decimation algorithm (SMD) that reduces the complexity of the mesh while still preserving the Laplacian eigenfunctions. Furthermore, Liu \etal~\cite{liu:2023:intrinsicsimplification} proposed an elegant solution for simplifying a shape based on intrinsic error metric (IEM), which also builds a map for extending scalar functions using geodesic barycentric coordinates. We perform the evaluation on the SHREC19 dataset~\cite{melzi:2019:shrec19}, remeshing the surfaces to 3k vertices with all the methods.


To achieve the desired scalability, the remeshing step in our pipeline must be fast enough to maintain the improvement obtained by reducing the size of the eigenproblem. 
In this regard, \Cref{fig:shrec19-remesh-times} shows that our technique achieves better time performance than the other remeshing algorithms and the sampling strategy of SFM.
 In contrast, SMD and IEM are decimation algorithms that iteratively reduce the mesh size, making them orders of magnitude slower and unsuitable for large meshes. Moreover, SMD requires a pre-computation of the Laplacian eigenbasis on the original surface to guide the decimation process and preserve the spectrum, introducing additional time complexity.

\subsection{Benefits of remeshing}\label{sec:results:benefits}


\begin{figure}[t]
    \centering
    \includegraphics[width=\columnwidth]{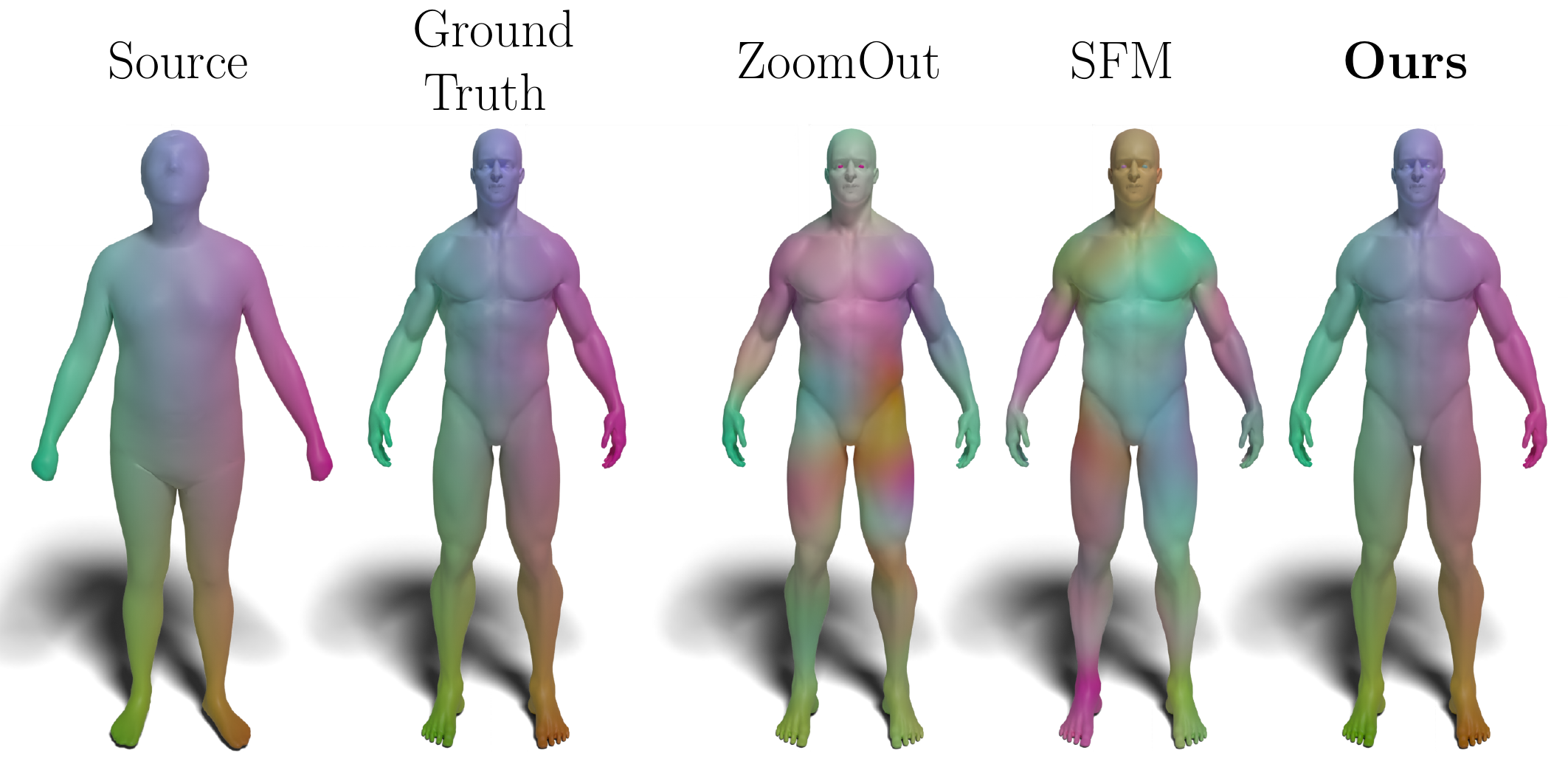}
    \includegraphics[width=\columnwidth]{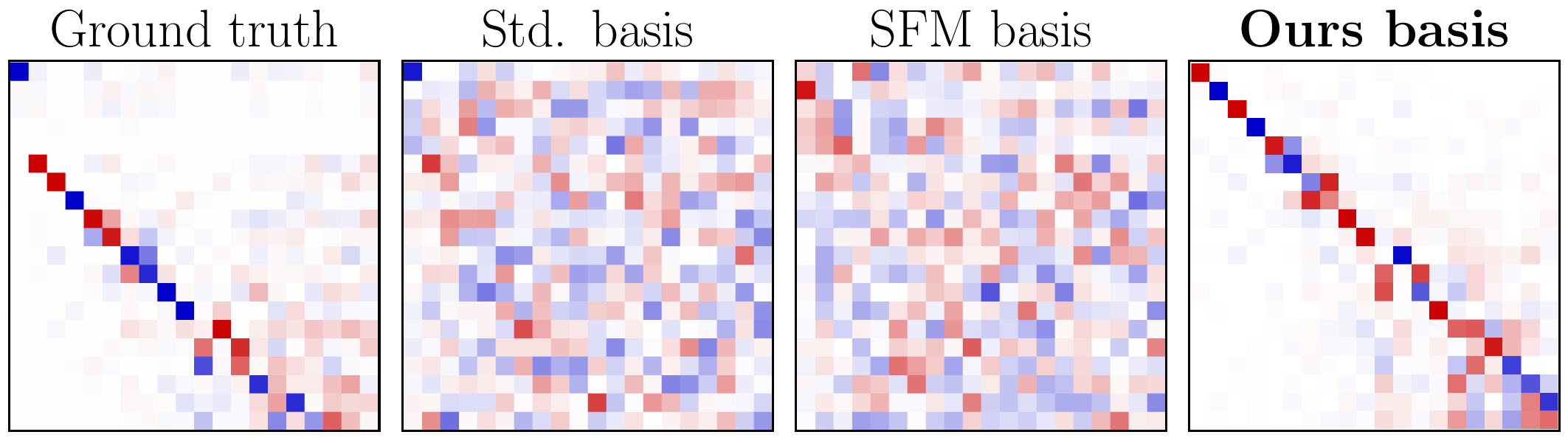}
    \caption{Top row: coordinate transfer between a pair from SHREC19 using ZoomOut on full-resolution meshes, SFM and our approach. Bottom row: The ground truth $20 \times 20$ functional map, compared with the alignment computed with FMaps~\cite{nogneng:2017:descriptorpreserve} from the full Laplacian eigenbasis, the SFM extended basis and our extended basis.}
    \label{fig:sfm-broken-map}
\end{figure}

In \Cref{sec:results:quality}, we discussed how we remove small disconnected components after the remeshing step. The example in \Cref{fig:sfm-broken-map} shows a comparison of mapping between a low-resolution mesh (Source), with less than 7k vertices, to a detailed mesh where eyes are represented with two disconnected components, each composed of 462 vertices out of the 27k of the entire mesh ($\sim$2\% of the total vertices). This is enough geometry to attract energy at lower frequencies, and when we try to align the first 20 eigenfunctions with FMaps~\cite{nogneng:2017:descriptorpreserve}, the resulting functional map turns out to be mostly noise. Initializing ZoomOut with such a map leads to meaningless correspondence, and SFM does not mitigate this issue, as it tries to preserve the original spectrum as much as possible. In contrast, by removing these components and mapping them to their closest surface points, we introduce a small error in the final correspondence, but at the same time, we ease the alignment of the initial eigenbases, resulting in more accurate and meaningful mapping. This is evident in the bottom row of \Cref{fig:sfm-broken-map}, where the ground truth functional map presents a gap in the alignment of the eigenfunctions due to the impossibility of mapping the disconnected eyes of the target shape in any point of the source shape. Using the real eigenfunctions or a close approximation yields an intense noise in the map, producing the meaningless mapping of ZoomOut and SFM. Instead, by mapping the eyes to their nearest point on the head, we can meaningfully align our extended basis.

\section{Conclusions}\label{sec:conclusions}

We presented a new pipeline for scalable shape matching that relies on a low-resolution representation to compute a functional map between dense shapes efficiently. A core part of this pipeline is our novel remeshing algorithm, which efficiently computes an intrinsic Delaunay triangulation of uniform surface sampling. This procedure is applied to reduce the computational cost of the functional maps framework and make it efficient also on very dense meshes. Furthermore, we use a fast and intuitive technique for extending scalar functions from low-resolution meshes to their dense counterparts, extending the computed correspondence to the original very high-resolution shapes. Our experimental evaluation proved that our procedure is effective on various types of shapes at different resolutions, outperforming other state-of-the-art solutions and solving topological issues related to the alignment of the Laplacian eigenbasis.

While using a triangulated mesh as a low-resolution representation could, in principle, be exploited with other types of shape-matching pipelines, we only tested it within the functional maps framework, using the Laplace-Beltrami eigenbasis or bases derived from it. Studying the behaviour of our method with different ``backends'', as well as exploring the relationships between the low-resolution and the high-resolution correspondences, would be an interesting matter for future investigations.

\bibliographystyle{eg-alpha-doi}
\bibliography{biblio}

\clearpage

\appendix

\section{Proof of Theorem \ref{thm:flatunion}}\label{sec:thmproof}
\begin{customthm}{2}
    Let $(\M, d_{\M})$ be a 2-dimensional metric space. If a general VD $\{P_i\}_{i=1}^{s}$ of $\M$ satisfies the \emph{FUP}, it also satisfies the closed ball property.
    \begin{proof}
        The first condition of the closed ball property trivially holds.

        Suppose $P_i \cap P_j$ is not empty, but it is not a closed 1-ball. Since $P_i \cup P_j$ is topologically flat, if their boundary is formed by two or more connected components, none of these can be a closed loop. But if the boundary between $P_i$ and $P_j$ is formed by two or more topological segments, then $P_i \cup P_j$ must contain a hole, which contradicts the second condition of the flat union property. Conversely, if their boundary is formed by a single connected component which is not a topological segment, then the boundary must form a loop inside $P_i \cup P_j$. Such loop would either contain $P_i$ or $P_j$, forming a hole in the other texel and violating the first condition of the flat union property.
        
        Suppose $P_i \cap P_j \cap P_k$ is not empty, but it is not a closed 0-ball. Every connected component of this intersection cannot be more than zero dimensional, so $P_i$, $P_j$, and $P_k$ must meet at more than one point. Since we proved that $P_i \cap P_j$ is a closed 1-ball, then $P_i \cap P_j \cap P_k$ must be formed by two points which are also the endpoints of $P_i \cap P_j$. A similar argument can be made for $P_j \cap P_k$ and $P_k \cap P_i$. Since $P_i \cup P_j \cup P_k$ is a closed 2-ball that contains three curves incident on the same two points, then two of these curves must form a closed loop containing the other one. Without loss of generality, suppose $P_i \cap P_j$ and $P_i \cap P_k$ form a closed loop. Since $P_i$ is a closed 2-ball, then the region enclosed by this loop must be $P_i$. But this region also contains the boundary between $P_j$ and $P_k$, which contradicts the fact that $P_i$, $P_j$, and $P_k$ are Voronoi texels. Hence, $P_i \cap P_j \cap P_k$ must be a closed 0-ball.
    \end{proof}
\end{customthm}

\section{Proof of Proposition \ref{prp:2ball}}\label{sec:prpproof}
\begin{customprp}{2}
    Let $\M = (V, E, F)$ be a manifold polygonal mesh. The mesh $\M$ is a closed 2-ball if and only if its Euler characteristic $\chi = |V| - |E| + |F|$ is equal to 1.
    \begin{proof}
        The Euler characteristic obeys to the equation $\chi = 2 - 2g - B$, where $g$ is the genus of the surface and $B$ is the number of boundary components. We also know that $\M$ is a closed 2-ball if and only if $g = 0$ and $B = 1$, thus if $\M$ is a closed 2-ball, we have $\chi = 1$.
        
        Suppose $\chi = 2 - 2g - B = 1$, then $B = 1 - 2g$, where $B, g$ are non-negative integers. If $g = 0$, then $B = 1$, and $\M$ is a closed 2-ball. Since $g > 0$ implies $B < 0$, there cannot be other valid solutions, and hence $\chi = 1$ implies that $\M$ is a closed 2-ball.
    \end{proof}
\end{customprp}

\section{Handling large triangles}\label{sec:resampling}
\begin{figure}[t]
    \centering
    \definecolor{myred}{rgb}{0.9,0,0}%
\definecolor{mygreen}{rgb}{0,0.9,0}%
\begin{tabular}{cc}
\begin{tikzpicture}[thick,scale=1.7, every node/.style={scale=1.4}]
    \coordinate (v0) at (0, 0);
    \coordinate (v1) at (1, 0.1);
    \coordinate (v2) at (0.4, 0.866);
    \filldraw[black] (v0) circle (2pt) node[anchor=east]{$i$};
    \filldraw[black] (v1) circle (2pt) node[anchor=west]{$j$};
    \filldraw[black] (v2) circle (2pt) node[anchor=south]{$k$};

    \draw (v0) -- (v1) -- (v2) -- cycle;
\end{tikzpicture}
&
\begin{tikzpicture}[thick,scale=1.7, every node/.style={scale=1.4}]
    \coordinate (v0) at (0, 0);
    \coordinate (v1) at (1, 0.5);
    \coordinate (v2) at (0.5, 1.866);
    \coordinate (n1) at (0.25, 0.933);

    \draw (v0) -- (v1) -- (v2);
    \draw[myred] (v0) -- (v2);
    \draw[mygreen] (n1) -- (v1);

    \filldraw[myred] (n1) circle (2pt) node[anchor=east]{$m_j$};
    \filldraw[black] (v0) circle (2pt) node[anchor=east]{$i$};
    \filldraw[black] (v1) circle (2pt) node[anchor=west]{$j$};
    \filldraw[black] (v2) circle (2pt) node[anchor=south]{$k$};
\end{tikzpicture}
\\
\begin{tikzpicture}[thick,scale=1.7, every node/.style={scale=1.4}]
    \coordinate (v0) at (0, 0);
    \coordinate (v1) at (1.5, 0.33);
    \coordinate (v2) at (0.75, 1.866);
    \coordinate (n0) at (1.125, 1.098);
    \coordinate (n1) at (0.375, 0.933);

    \draw (v0) -- (v1);
    \draw[myred] (v0) -- (v2);
    \draw[myred] (v1) -- (v2);
    \draw[mygreen] (n0) -- (n1);
    \draw[mygreen] (n1) -- (v1);
    \filldraw[myred] (n0) circle (2pt) node[anchor=west]{$m_i$};
    \filldraw[myred] (n1) circle (2pt) node[anchor=east]{$m_j$};
    \filldraw[black] (v0) circle (2pt) node[anchor=east]{$i$};
    \filldraw[black] (v1) circle (2pt) node[anchor=west]{$j$};
    \filldraw[black] (v2) circle (2pt) node[anchor=south]{$k$};
\end{tikzpicture}
&
\begin{tikzpicture}[thick,scale=1.7, every node/.style={scale=1.4}]
    \coordinate (v0) at (0, 0);
    \coordinate (v1) at (2, 0);
    \coordinate (v2) at (1.3, 2*0.866);

    \coordinate (n0) at (1.65, 0.866);
    \coordinate (n1) at (0.65, 0.866);
    \coordinate (n2) at (1, 0);

    \draw[myred] (v0) -- (v1) -- (v2) -- cycle;
    \draw[mygreen] (n0) -- (n1) -- (n2) -- cycle;
    
    \filldraw[black] (v0) circle (2pt) node[anchor=east]{$i$};
    \filldraw[black] (v1) circle (2pt) node[anchor=west]{$j$};
    \filldraw[black] (v2) circle (2pt) node[anchor=south]{$k$};
    \filldraw[myred] (n0) circle (2pt) node[anchor=west]{$m_i$};
    \filldraw[myred] (n1) circle (2pt) node[anchor=east]{$m_j$};
    \filldraw[myred] (n2) circle (2pt) node[anchor=north]{$m_k$};
\end{tikzpicture}
\end{tabular}
    \caption{Resampling scheme of the triangles, depending on the number of split edges. Red: long edges, to be split. Green: added connectivity.}
    \label{fig:resampling-tris}
\end{figure}
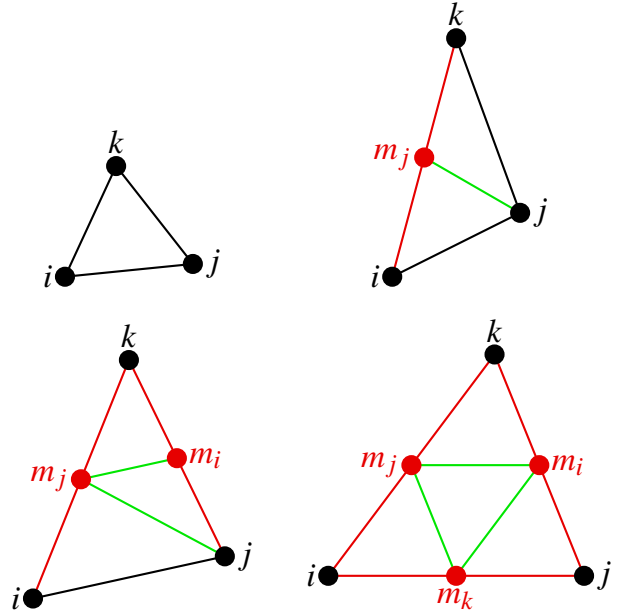

In most cases, we wish to preserve any large planar faces appearing in the original mesh, as they represent large planar regions which should be preserved in the final triangulation. However, sometimes these kind of faces could introduce distorted triangles and degenerate angles, and thus it becomes preferable to introduce extra triangulation and removing them. For this task, we introduce a resampling strategy which can be performed as a preprocessing step. 

Statistically, triangular meshes have triangle count $\sim$2 times the number of vertices. Given the total area $A_{\M}$ of the original shape $\M$, we can estimate the average triangle area for the output mesh as $A_E \approx \nicefrac{A_{\M}}{2 s}$, with $s$ the number of vertices of the remeshing. An equilateral triangle of area $A_E$ has each side of length $\rho = \sqrt{\nicefrac{2 A_E}{\sqrt{3}}}$, so we split in half every edge with length greater than $\rho$ and we insert a new vertex in the midpoint.

We then split mesh triangles depending on the number of incident split edges. If the triangle contains no split edges, we leave it as is. In case of a single split edge, we connect the midpoint to the opposite vertex, forming two triangles. When we have two split edges, we connect the midpoints to form a triangle and a quad. Then, to mitigate the formation of ill-shaped triangles, we divide the quad connecting the opposite vertices with the largest angle sum. Finally, if all its edges are split, we connect the midpoints to form four triangles. The process can then be iterated until no edges are oversize. The four cases for splitting a triangle are depicted in \Cref{fig:resampling-tris}.

\section{BadTOSCA dataset}\label{sec:badtosca}

\begin{figure*}[t!]
    \centering
    \includegraphics[width=0.65\columnwidth]{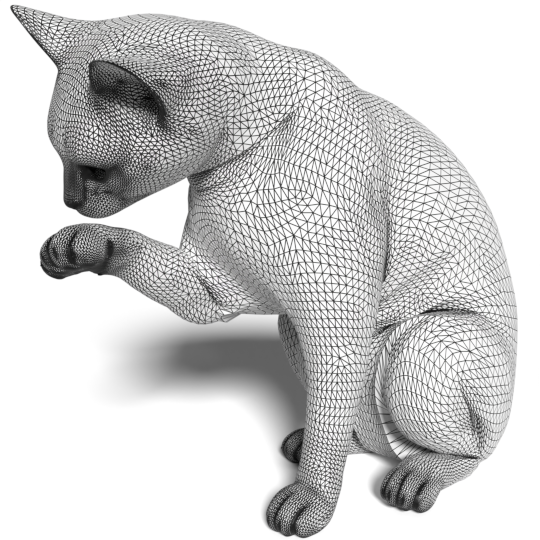}%
    \includegraphics[width=0.65\columnwidth]{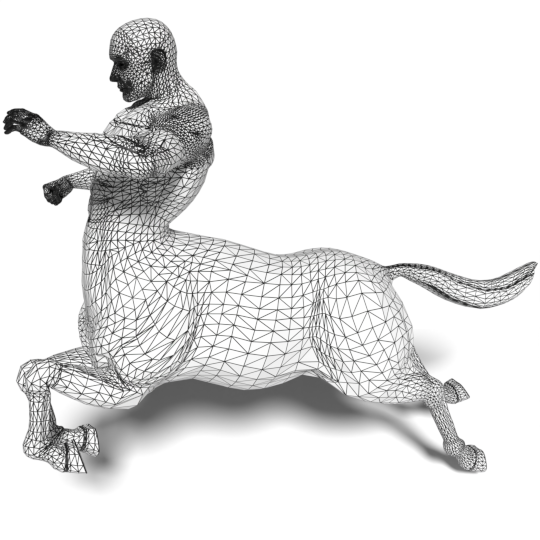}%
    \includegraphics[width=0.65\columnwidth]{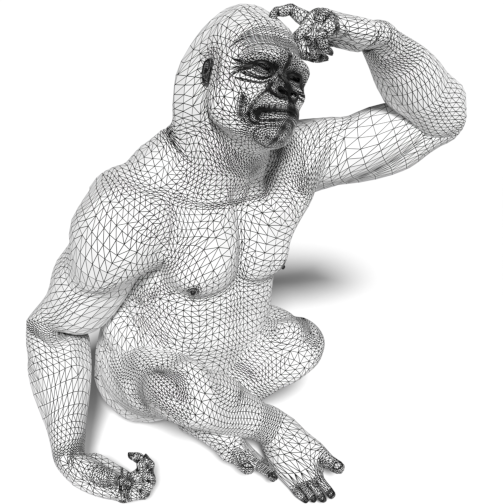}
    \includegraphics[width=0.65\columnwidth]{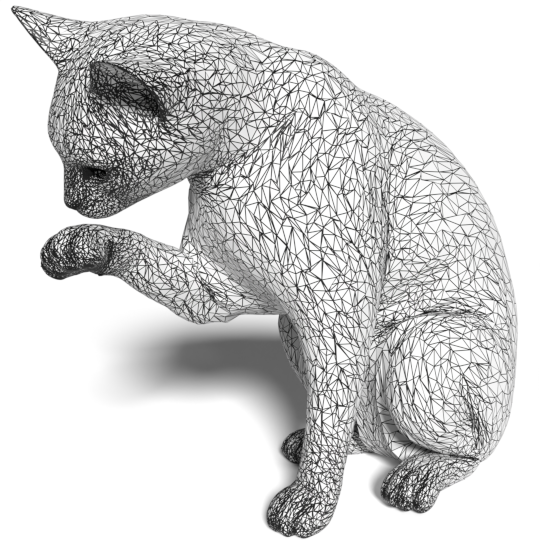}%
    \includegraphics[width=0.65\columnwidth]{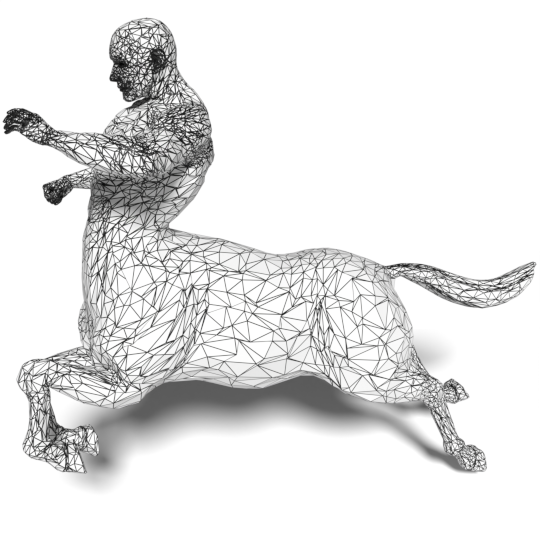}%
    \includegraphics[width=0.65\columnwidth]{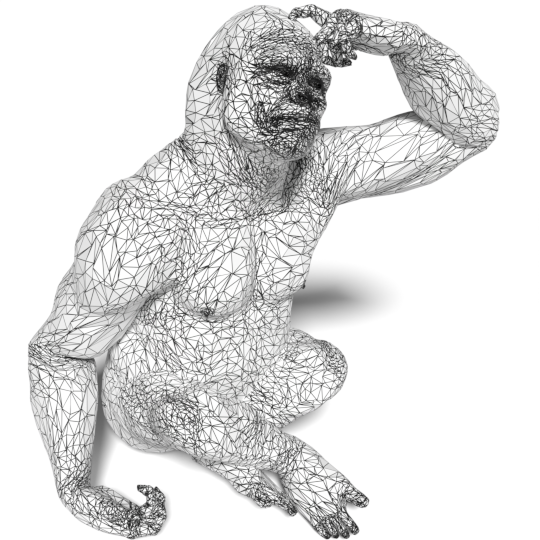}
    \caption{Meshes from the TOSCA dataset and their altered counterparts in BadTOSCA.}
    \label{fig:badtosca-examples}
\end{figure*}

In the main manuscript, we introduced a new dataset for testing our algorithm, which we referred to as BadTOSCA. This dataset aims to introduce additional challenge to the well-known TOSCA dataset~\cite{bronstein:2008:tosca} by altering the geometry and breaking the strong isometry between pairs of the same class. We recall the structure of the TOSCA dataset:
\begin{itemize}
    \item the dataset is subdivided into $k$ collections $C_1, \cdots, C_k$ of shapes;
    \item each class $C_i$ contains $s$ triangular meshes $\M_{i_1}, \cdots, \M_{i_s}$ representing non-rigid deformations of the same shape;
    \item every two shapes $\M_p, \M_q$ belonging to the same class $C_i$ are near perfectly isometric, have the same number of vertices, the same connectivity, and their correspondence $\Pi$ is the identity matrix.
\end{itemize}

To generate the dataset, we first alter every shape independently from all the others. Given a shape $\M = (V, E, T)$, we generate an altered shape $\Hat{\M} = (\Hat{V}, E, T)$ by moving the position of the vertices, but preserving the connectivity. This ensures that the overall geometry is left unchanged, but at the same time that the strong isometry, the bijective correspondence, and the isomorphic connectivity cannot be exploited (see \Cref{fig:badtosca-examples} for a reference).

For each vertex $v$, we compute the set $N_T(v)$ of triangles incident on $v$, and we select a random triangle $t \in N_T(v)$. We then compute random barycentric coordinates $\boldsymbol{\lambda}(v) = (\lambda_1(v), \lambda_2(v), \lambda_3(v))$, and place $v$ on triangle $t$ at $\boldsymbol{\lambda}(v)$. Said $\mathbf{V} \in \Realsn{|V| \times 3}$ the matrix of the vertex positions of $\M$ and $\hat{\mathbf{V}} \in \Realsn{|\hat{V}| \times 3}$ the matrix of vertex positions of $\hat{\M}$, we can then use the barycentric coordinates of the altered vertices to build a sparse matrix $\mathbf{U} \in \Realsn{|\hat{V}| \times |V|}$ such that $\hat{\mathbf{V}} = \mathbf{U}\mathbf{V}$.

Given two meshes $\M_i = (V_i, E, T), \M_j = (V_j, E, T)$ belonging to the same class, let $\hat{\M}_i = (\hat{V}_i, E, T)$ and $\hat{\M}_j = (\hat{V}_j, E, T)$ be the corresponding altered meshes, and let $\mathbf{U}_i$ and $\mathbf{U}_j$ be the mapping of the vertices. By construction, we know that $\hat{\mathbf{V}}_i = \mathbf{U}_i\mathbf{V}_i$. Furthermore, since $\M_i$ and $\M_j$ are near perfectly isometric, the product $\mathbf{U}_j\mathbf{V}_i$ gives us the same type of altering that generated $\hat{\M}_j$ from $\M_j$, but onto the geometry of $\M_i$. Thus, for building the mapping $\Pi_{ij}$ between $\hat{\M}_i$ and $\hat{\M}_j$ we apply a nearest neighbor search between $\mathbf{U}_i\mathbf{V}_i$ and $\mathbf{U}_j\mathbf{V}_i$.

\section{Closest surface point mapping}\label{sec:mapping}

Despite the simplicity of the method, projecting onto the closest surface point and using barycentric coordinates to interpolate values onto the original surface proves to be a valid solution, as shown by the results on the main manuscript.

\begin{figure*}[t]
    \centering
    \includegraphics[width=\textwidth]{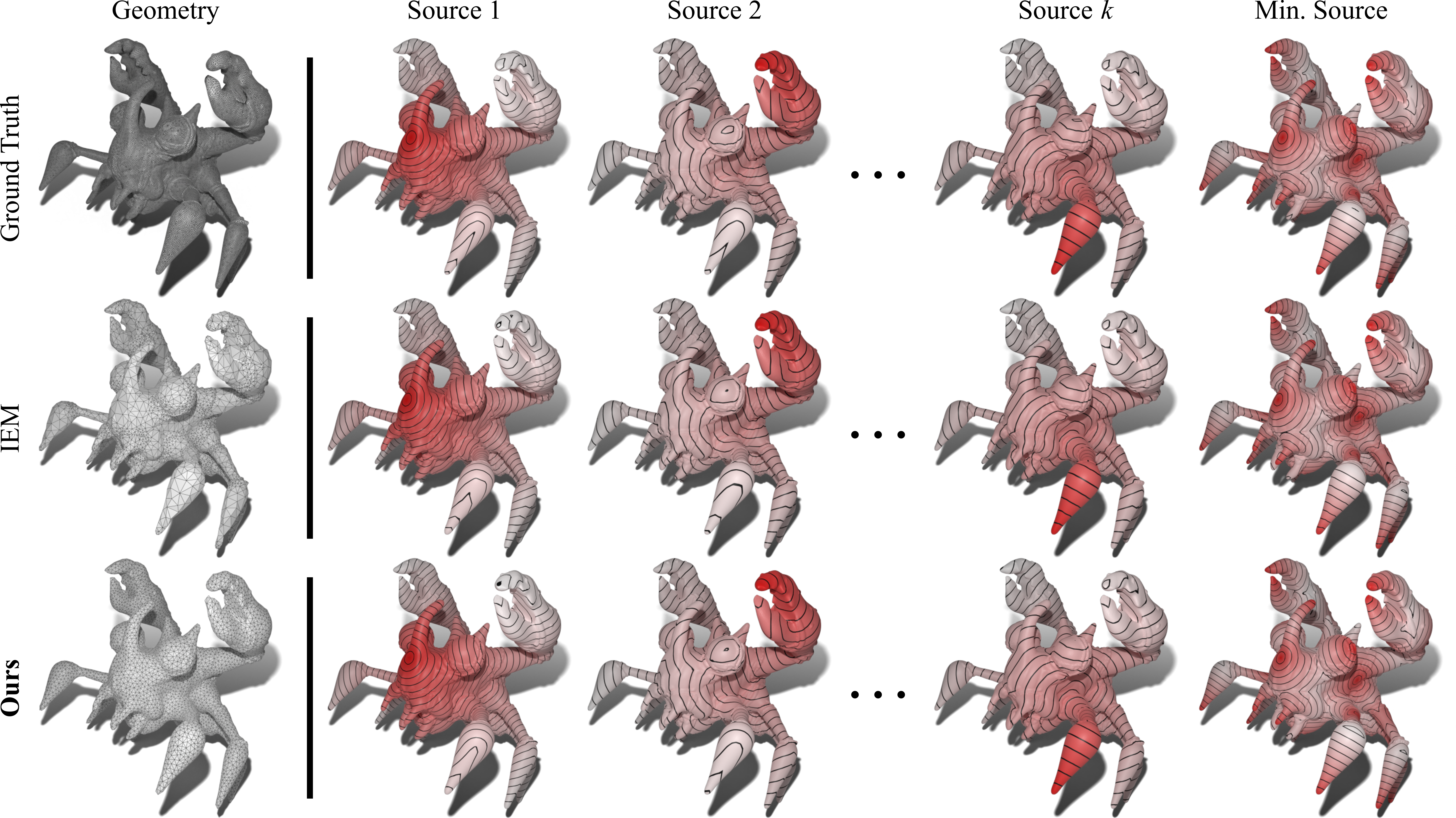}
    \caption{Comparison between our method (last row) and IEM~\cite{liu:2023:intrinsicsimplification} (second row) extending geodesic distances to the high-resolution surface. The top row shows the ground truth geodesics.}
    \label{fig:geodesics-compare}
\end{figure*}

We highlight that our method can compete with the intrinsic simplification method proposed by Liu \etal~\cite{liu:2023:intrinsicsimplification} also in other settings. \Cref{fig:geodesics-compare} shows an example of how geodesic distances can be computed in the low-resolution surface (10k vertices for both methods, shown in the first column) and then extended to the high-resolution mesh ($\sim$120k vertices). For this experiment, we sample $k = 30$ source points $p_1,\dots,p_k$ from the surface and compute the exact geodesic distances $d_{\mathrm{gt}}(p_i, v)$ from each source $p_i$ to every vertex $v$ of the surface, also considering the minimum distance from the sample set $\min_{p_i}(d_{\mathrm{gt}}(p_i, v))$ (last column). We then do the same for the remeshed surface, using as source the points closest to the original samples, and extending the distance functions to the full-resolution shape using the barycentric coordinates of the closest surface point in our case, and the approximated geodesic barycentric coordinates for IEM, respectively producing approximated geodesic distances $d_{\mathrm{ours}}(p_i, v)$ and $d_{\mathrm{IEM}}(p_i, v)$.


\begin{table}[t]
    \centering
    \setlength{\tabcolsep}{4pt}
{\footnotesize%
\begin{tabular}{ccc}
    Point  & Ours  & IEM    \\
    \hline
    $p_{1}$         & 1.81    & 1.80    \\
    $p_{2}$         & 1.54    & 0.72    \\
    $p_{3}$         & 1.62    & 0.79    \\
    $p_{4}$         & 1.04    & 0.48    \\
    $p_{5}$         & 1.05    & 0.59    \\
    $p_{6}$         & 1.44    & 0.70    \\
    $p_{7}$         & 0.95    & 0.39    \\
    $p_{8}$         & 1.35    & 0.88    \\
    $p_{9}$         & 1.28    & 0.49    \\
    $p_{10}$        & 1.04    & 0.60    \\
    $p_{11}$        & 1.25    & 1.04    \\
    $p_{12}$        & 2.37    & 0.77    \\
    $p_{13}$        & 1.93    & 1.11    \\
    $p_{14}$        & 1.10    & 0.80    \\
    $p_{15}$        & 0.84    & 0.49    \\
    \phantom{x} & \phantom{x} & \phantom{x}
\end{tabular}%
\begin{tabular}{ccc}
    Point  & Ours  & IEM    \\
    \hline
    $p_{16}$        & 1.50    & 0.56    \\
    $p_{17}$        & 1.96    & 0.51    \\
    $p_{18}$        & 1.29    & 0.46    \\
    $p_{19}$        & 1.79    & 1.54    \\
    $p_{20}$        & 1.91    & 2.32    \\
    $p_{21}$        & 1.94    & 0.44    \\
    $p_{22}$        & 1.72    & 0.79    \\
    $p_{23}$        & 1.13    & 0.61    \\
    $p_{24}$        & 1.08    & 0.52    \\
    $p_{25}$        & 1.03    & 1.48    \\
    $p_{26}$        & 1.05    & 1.61    \\
    $p_{27}$        & 0.74    & 0.73    \\
    $p_{28}$        & 1.24    & 2.31    \\
    $p_{29}$        & 1.98    & 0.57    \\
    $p_{30}$        & 1.74    & 1.95    \\
    Min. dist       & 2.61    & 3.86    \\
\end{tabular}%
}
    \caption{Normalized error of approximated geodesics from each source point for both our method and IEM.}
    \label{tab:geodesics-compare}
\end{table}

For evaluating the error, for each $p_i$ we compute the differences
\begin{equation}
\begin{gathered}
    e_{\mathrm{ours}}(p_i, v)
    =
    \frac{
        d_{\mathrm{gt}}(p_i, v) - d_{\mathrm{ours}}(p_i, v)
    }{
        \max_{v}(d_{\mathrm{gt}}(p_i, v))
    }\,,
\\
    e_{\mathrm{intrinsic}}(p_i, v)
    =
    \frac{
        d_{\mathrm{gt}}(p_i, v) - d_{\mathrm{intrinsic}}(p_i, v)
    }{
        \max_{v}(d_{\mathrm{gt}}(p_i, v))
    }\,,
\end{gathered}
\end{equation}
which are normalized by maximum distance to $p_i$, to ensure that each function acts at the same scale. Then, we aggregate the error by computing the norms $\|e_{\mathrm{ours}}(p_i, \cdot)\|_{\M}$ and $\|e_{\mathrm{intrinsic}}(p_i, \cdot)\|_{\M}$, where $\|f\|_{\M}^2 = \int_{\M} f^2(x)\ \mathrm{d}x$.

In contrast to our method, IEM produces a mapping that approximates geodesic barycentric coordinates during the simplification. However, the geodesic paths between adjacent vertices in the remeshed surface are deformed to straight lines, negatively affecting the approximation of the full-resolution geodesics and producing results that are comparable to ours. The results from \Cref{tab:geodesics-compare} shows that, while IEM obtains better results most of the time, the error is on the same scale as our solution. Furthermore, while IEM took 25.796 seconds for remeshing the surface and producing the mapping, our algorithm remeshed the surface in 1.109 seconds and produced the map in 638 milliseconds, totalling 1.747 seconds and achieving a 14.75 speedup.

\end{document}